\documentclass[onecolumn]{IEEEtran}

\usepackage{epsfig,amsmath,amssymb,amsbsy,amsthm,epsf,amsthm,scalefnt,subfig,multirow,algorithm,algorithmic,mathtools}
\usepackage{xcolor}
\usepackage{float}
\usepackage{cite}
\usepackage[T1]{fontenc}
\usepackage[doublespacing]{setspace}

\def\inf{\mathop{\mathrm{inf}}}

\def\b0{{\pmb{0}}} 

\newcommand\ignore[1]{}

\DeclareMathOperator*{\esssup}{ess\,sup}
\DeclareMathOperator*{\essinf}{ess\,inf}

\begin{document}

\title{Real-Time Detection of Hybrid and Stealthy Cyber-Attacks in Smart Grid}

\author{Mehmet Necip~Kurt,~
        Yasin~Y{\i}lmaz,~\IEEEmembership{Member,~IEEE},~
        and~~Xiaodong~Wang,~\IEEEmembership{Fellow,~IEEE}
\thanks{This work was supported in part by the U.S. National
Science Foundation (NSF) under Grant ECCS-1405327, and in part by the
U.S. Office of Naval Research under Grant N000141410667. The work of
Y. Y{\i}lmaz was supported in part by the NSF under Grant CNS-1737598 and
in part by the Southeastern Center for Electrical Engineering Education.}%
\thanks{M.\,N. Kurt and X. Wang are with the Department
of Electrical Engineering, Columbia University, New York, NY 10027, USA (e-mail: m.n.kurt@columbia.edu; wangx@ee.columbia.edu).}
\thanks{Y. Y{\i}lmaz is with the Department of Electrical Engineering, University of South Florida, Tampa, FL 33620, USA (e-mail: yasiny@usf.edu).}\vspace{-2.5ex}}

\maketitle

\begin{abstract}
For a safe and reliable operation of the smart grid, timely detection of cyber-attacks is of critical importance. Moreover, considering smarter and more capable attackers, robust detection mechanisms are needed against a diverse range of cyber-attacks. With these purposes, we propose a robust online detection algorithm for (possibly combined) false data injection (FDI) and jamming attacks, that also provides online estimates of the unknown and time-varying attack parameters and recovered state estimates. Further, considering smarter attackers that are capable of designing stealthy attacks to prevent the detection or to increase the detection delay of the proposed algorithm, we propose additional countermeasures. Numerical studies illustrate the quick and reliable response of the proposed detection mechanisms against hybrid and stealthy cyber-attacks. \vspace{-0.3cm}
\end{abstract}


\section{Introduction}

\subsection{A Brief Overview of Cyber-Attacks and Countermeasures in Smart Grid}

Due to the integration of advanced signal processing, communication, and control technologies, smart grid relies on a critical cyber infrastructure that is subject to adversarial cyber threats \cite{HHe16,Liang16,wang2013cyber,yan2012survey}. The smart grid is regulated based on estimated system states and the main aim of attackers is to damage/mislead the state estimation mechanism and thereby to cause wrong/manipulated decisions in the energy management system of the smart grid. Some potential consequences of a successful cyber-attack are regional power blackouts, {manipulated electricity market prices \cite{Xie10,moslemi2017}, and destabilization of the power grid \cite{ayar2017}.} Such cyber-attacks are also seen in practice. For instance, on December 23, 2015, the Ukrainian power system was attacked and the resulting power blackout affected around 200,000 people for several hours \cite{GLiang17}.

The Ukraine attack has demonstrated that attackers have more capabilities than predicted \cite{GLiang17}. Namely, (i) attackers can access and monitor the power system over long periods of time without being detected, (ii) attackers are able to perform cyber-attacks by hacking smart grid components (smart meters, control centers, etc.), manipulating/jamming the network communication channels, and accessing and manipulating the database of the control center \cite{GLiang17,Amin09,Liang16}. Hence, cyber-attacks significantly threaten the safe and reliable operation of the power grid in practice. Effective countermeasures need to be developed considering the worst-case scenarios where the attackers are fully capable of performing a diverse range of cyber-attacks. The first step in a defense mechanism is early detection of cyber-attacks. After detecting an attack, effective mitigation schemes should then be implemented.

Recently, the false data injection (FDI) attacks \cite{Liang16,Liu09,Tan17,moslemi2017fast} and the jamming attacks \cite{Amin09,Sargolzaei17,YLi15,Deka15} against the smart grid are extensively studied in the literature and several detectors are proposed. The proposed detectors are mostly outlier detectors, i.e., they classify a sample measurement as either normal or anomalous. Conventional detectors classify a measurement as anomalous if the measurement residual exceeds a certain threshold \cite{Abur04,Liu09,Manandhar14,Brumback87,Rawat15}. More advanced machine learning techniques are also considered for classification of anomalous measurements \cite{Esmalifalak17,Ozay16}. {Moreover, in \cite{moslemi2017fast}, firstly a Markov graph model for system states is learned under normal system operation and then attacks/anomalies are detected based on the consistency of new measurements compared to the learned nominal model. Further, in \cite{bretas2017}, based on the least squares (LS) state estimator, a multi-step procedure is presented to detect and classify cyber-attacks on meter measurements, network line parameters, and network topology, and then to make corrections for attack mitigation.}

{In \cite{zhao2017robust,zhao2017attack,gandhi2010robust}, robust extended Kalman filters have been proposed where the main aim is to bound the effects of outliers on the state estimation mechanism. No specific attack types are considered so that using such schemes, it is not possible to distinguish a real attack from random outliers, e.g., due to heavy-tailed non-Gaussian noise processes. Moreover, such schemes have breakdown points such that if outliers, significantly far away from the nominal measurements, are observed, then the proposed filters fail to keep track of the system state.}

In order to improve the time resolution and also to detect cyber-attacks more reliably, several online detectors based on the quickest detection theory are proposed. For instance, in \cite{Li_15} and \cite{Huang16}, cumulative sum (CUSUM)-based schemes are considered to detect FDI attacks where the state estimation is based on the conventional LS methods. More recently, in \cite{Necip18}, CUSUM-based detection schemes are proposed to detect FDI and denial of service (DoS) attacks (separately) in a dynamic setting and their advantages over the outlier detectors and the LS-based detectors are demonstrated. {Further, in \cite{yang2016false}, a nonparametric CUSUM detector is proposed that do not assume any attack model and only evaluates the deviation of meter measurements from the baseline statistics, i.e., normal system operation. In \cite{sun2013anomaly}, a window-based CUSUM detector is proposed for detection of FDI attacks where the attack parameters of interest are estimated based on the most recent sliding window of measurements.}

\subsection{Contributions}

In this paper, we propose robust mechanisms for timely detection of potentially combined and stealthily designed FDI and jamming attacks. The proposed mechanisms are tightly connected to an estimation mechanism, which makes both the detection and state estimation schemes robust against unknown and time-varying attack variables. In particular, online maximum likelihood estimates (MLEs) of the attack types, set of attacked meters, and the attack magnitudes are used in attack detection. Moreover, recovered state estimates are computed based on the online MLE estimates of the attack variables. No restrictive assumptions are made about an attacker's strategy, i.e., an attacker can design and perform arbitrarily combined FDI and jamming attacks, targeting any subset of meters in any magnitude and can also change its attack parameters over time. Further, considering the possibility of smarter and more capable attackers, additional countermeasures are proposed against stealthily designed cyber-attacks. These make the proposed detection schemes highly robust against a  significantly wide range of potential cyber-attacks targeting the smart grid.

Since the smart grid is a highly complex network, any anomaly/failure in a part of the system can quickly spread over the network and lead to new unpredicted failures. Hence, timely attack detection and mitigation is crucial. In this paper, for timely detection, we present real-time detection mechanisms. Moreover, to help for timely attack mitigation and quick system recovery, we provide online estimates of the attack types, set of attacked meters and attack magnitudes. Note that having an estimate for the attack type can be useful since different countermeasures may need to be employed against different types of attacks. Further, considering that the real power grid is a huge network consisting of many meters, an estimate of the attacked meters can be critical for a timely and effective attack mitigation, e.g., via isolating the attacked meters during the recovery procedure. Moreover, estimates of attack magnitudes are needed to recover attack-free states.

We list our main contributions as follows:

\begin{itemize}
\item A novel low-complexity online detection and estimation algorithm is proposed against (possibly) combined FDI and jamming attacks. The proposed algorithm is robust to unknown and time-varying attack types, magnitudes, and set of attacked meters. Further, recovered state estimates and closed-form online MLE estimates of the attack variables are presented.
\item Stealthy attacks against CUSUM-based detectors and particularly against the proposed algorithm are introduced and analyzed.
\item Several countermeasures are proposed against the considered stealthy attacks.
\end{itemize}

\subsection{Organization}

The remainder of the paper is organized as follows. In Sec.~\ref{sec:system}, the system model, attack models, state estimation mechanism, and the problem formulation are presented. In Sec.~\ref{sec:combined}, an online cyber-attack detection and estimation algorithm is presented. In Sec.~\ref{sec:stealth}, stealthy attacks against CUSUM-based detectors are introduced and analyzed. Also, countermeasures against the considered stealthy attacks are presented. In Sec.~\ref{sec:numerical}, the proposed detection schemes are evaluated extensively via simulations. Finally, the paper is concluded in Sec.~\ref{sec:conc}. Boldface letters denote vectors and matrices, and all vectors are column vectors.

\section{System Model and Problem Formulation} \label{sec:system}

\subsection{System Model}

The actual power grid is regulated based on a nonlinear AC power flow model \cite{Liang16}. On the other hand, the approximate linearized (around an operating point) DC power flow model is a good approximation that is widely used in the literature to describe the operation of the power grid \cite{Abur04,Liu09,Cui12}. Furthermore, static system model and consequently conventional static (LS) state estimation are not effective in capturing the dynamics of a power system due to time-varying load and power generation \cite{Tan17}. In addition, attack detection mechanisms based on static estimators are not effective in detecting time-varying cyber-attacks and {structured ``stealth'' FDI attacks \cite{Liu09}, for which dynamic state estimator-based detectors are known to be effective \cite{Necip18,Zhao17}.}

We then model the power grid, consisting of $N+1$ buses and $K$ meters, as a discrete-time linear dynamic system based on the commonly employed linear DC model \cite{Abur04,Liu09,Cui12} as follows:
\begin{gather} \label{eq:state_upd}
\mathbf{x}_t = \mathbf{A} \mathbf{x}_{t-1} + \mathbf{v}_t, \\ \label{eq:meas_model}
\mathbf{y}_t = \mathbf{H} \mathbf{x}_t + \mathbf{w}_t,
\end{gather}
where $\mathbf{x}_{t} = [x_{1,t}, x_{2,t}, \dots, x_{N,t}]^\mathrm{T}$ is the state vector denoting the phase angles of $N$ buses (one of the buses is considered as a reference bus), $\mathbf{A} \in \mathbb{R}^{N \times N}$ is the state transition matrix, $\mathbf{v}_t = [v_{1,t}, v_{2,t}, \dots, v_{N,t}]^\mathrm{T} \sim \mathbf{\mathcal{N}}(\mathbf{0},\sigma_v^2 \, \mathbf{I}_N)$ is the process noise vector, $\mathbf{I}_N$ is an ${N \times N}$ identity matrix, and $\cdot^\mathrm{T}$ is the transpose operator. Further, $\mathbf{y}_t = [\mathbf{y}_{1,t}^\mathrm{T}, \mathbf{y}_{2,t}^\mathrm{T}, \dots, \mathbf{y}_{K,t}^\mathrm{T}]^\mathrm{T}$ is the vector consisting of meter measurements, $\mathbf{y}_{k,t} = [y_{k,t,1}, y_{k,t,2}, \dots, y_{k,t,\lambda}]^\mathrm{T}$ is the measurement vector for meter $k$, $\mathbf{H} \in \mathbb{R}^{K \lambda \times N}$ is the measurement matrix, $\mathbf{w}_t = [\mathbf{w}_{1,t}^\mathrm{T}, \mathbf{w}_{2,t}^\mathrm{T}, \dots, \mathbf{w}_{K,t}^\mathrm{T}]^\mathrm{T} \sim \mathbf{\mathcal{N}}(\mathbf{0},\sigma_w^2 \, \mathbf{I}_{K \lambda})$ is the measurement noise vector, and $\mathbf{w}_{k,t} = [w_{k,t,1}, w_{k,t,2}, \dots, w_{k,t,\lambda}]^\mathrm{T}$ is the measurement noise vector for meter $k$. Note that in each time interval between $t-1$ and $t$, $\lambda \in \{1,2,3,\dots\}$ measurements are taken at each meter, where $\lambda$ is usually small, and the collected measurements between $t-1$ and $t$ are processed at time $t$. To increase the measurement redundancy against noise and also to estimate the unknown attack parameters more reliably in case of a cyber-attack, $\lambda$ needs to be chosen higher. 

{In general, the state transition and measurement matrices can also be dynamic. For instance, due to changes in network topology, i.e., on and off states of the switches and line breakers in the power grid, the measurement matrix may vary over time. In that case, instead of modeling the smart grid as a linear time-invariant system as in \eqref{eq:state_upd} and \eqref{eq:meas_model}, we can model it as a linear time-varying system where we can replace $\mathbf{A}$ and $\mathbf{H}$ by $\mathbf{A}_t$ and $\mathbf{H}_t$, respectively. The results presented in this study can be generalized to the case of linear time-varying system model as long as $\mathbf{A}_t$ and $\mathbf{H}_t$ are known by the system controller at each time $t$.}

\subsection{Attack Models}

We assume that at an unknown time $\tau$, a cyber-attack is launched to the system, where we particularly consider FDI attacks, jamming attacks, and their combination. The attack types, attack magnitudes, and the set of attacked meters can be time-varying. But, during a time interval, i.e., between $t-1$ and $t$, we assume that the attack parameters stay constant. Next, we explain the attack models under consideration.

\subsubsection{FDI Attack}

In case of an FDI attack, additive malicious data are injected into the measurements of a subset of meters. In practice, an FDI attack can be performed by manipulating the network communication channels or hacking meters and/or control centers in the smart grid \cite{Liang16,GLiang17}. The measurement model in case of an FDI attack takes the following form:
\begin{gather} \label{eq:fdi_model}
\mathbf{y}_t = \mathbf{H} \mathbf{x}_t + \mathbf{a}_t + \mathbf{w}_t, ~~ t \geq \tau,
\end{gather}
where $\mathbf{a}_t = [\mathbf{a}_{1,t}^\mathrm{T}, \mathbf{a}_{2,t}^\mathrm{T}, \dots, \mathbf{a}_{K,t}^\mathrm{T}]^\mathrm{T}$ denotes the injected false data at time $t$. Since the attack magnitudes are assumed to be constant between $t-1$ and $t$, for meter $k$, $\mathbf{a}_{k,t} = \mathbf{1}_{\lambda \times 1} \, \mathrm{a}_{k,t}$, where $\mathbf{1}_{\lambda \times 1}$ is a $\lambda \times 1$ vector consisting of $1$s. Note that if meter $k$ is not under an FDI attack at time $t$, then $a_{k,t} = 0$, otherwise $a_{k,t} \neq 0$.

\subsubsection{Jamming Attack}

In case of a jamming attack, we assume that the attacker constantly emits additive white Gaussian noise (AWGN) to the network communication channels to compromise a subset of meter measurements. We consider jamming with AWGN since (i) it is a commonly employed jamming model in the literature \cite{JGao15,Gezici16}, (ii) it is a simple attacking strategy to perform, and (iii) in an additive noise channel with Gaussian input, for a given mean and variance, among all noise distributions, the Gaussian noise maximizes the mean squared error of estimating the channel input given the channel output \cite{Kay93,Gezici16}. Hence, an attacker can jam the communication channels with AWGN to maximize its damage on the state estimation mechanism.

In case of a jamming attack, the measurement model can be written as follows:
\begin{gather} \label{eq:jamming_model}
\mathbf{y}_t = \mathbf{H} \mathbf{x}_t + \mathbf{w}_t + \mathbf{n}_t, ~~ t \geq \tau,
\end{gather}
where $\mathbf{n}_t = [\mathbf{n}_{1,t}^\mathrm{T}, \mathbf{n}_{2,t}^\mathrm{T}, \dots, \mathbf{n}_{K,t}^\mathrm{T}]^\mathrm{T} \sim \mathbf{\mathcal{N}}(\mathbf{0}, \mathrm{diag}(\pmb{\sigma}_t))$ denotes the jamming noise, $\pmb{\sigma}_t = [\pmb{\sigma}_{1,t}^\mathrm{T}, \pmb{\sigma}_{2,t}^\mathrm{T}, \dots, \pmb{\sigma}_{K,t}^\mathrm{T}]^\mathrm{T}$, and $\pmb{\sigma}_{k,t} = \mathbf{1}_{\lambda \times 1} \, \sigma_{k,t}^2$ where $\sigma_{k,t}^2$ is the variance of the jamming noise targeting meter $k$ at time $t$. If meter $k$ is not under a jamming attack at time $t$, then $\sigma_{k,t}^2 = 0$, otherwise $\sigma_{k,t}^2 > 0$.

\subsubsection{Hybrid Attack}

In case of a hybrid (combined) attack, FDI and jamming attacks are simultaneously launched to the system and hence the measurement model takes the following form:
\begin{gather} \label{eq:meas_attacked}
\mathbf{y}_t = \mathbf{H} \mathbf{x}_t + \mathbf{a}_t + \mathbf{w}_t + \mathbf{n}_t, ~~ t \geq \tau.
\end{gather}
For meter $k$ under both FDI and jamming attacks at time $t$, $a_{k,t} \neq 0$ and $\sigma_{k,t}^2 > 0$. Since the FDI and jamming attacks can be considered as special cases of hybrid attacks, we consider \eqref{eq:meas_attacked} as the measurement model under the attacking regime, i.e., for $t \geq \tau$.


\textit{Remark 1:} If the noise terms in the normal system operation are AWGN (as in \eqref{eq:state_upd} and \eqref{eq:meas_model}) and the jamming noise terms are mutually independent over the meters, then the considered hybrid FDI/jamming attacks span all possible data attacks. This is due to the fact that a Gaussian random variable is defined by its mean and variance, and through the hybrid attacks, mean and variance of the density of meter measurements can be arbitrarily changed (cf. \eqref{eq:meas_attacked}). For instance, in case of a DoS attack, meter measurements are blocked and only a random or zero signal is received at the control center \cite{Amin09,Sargolzaei17,YLi15}. Hence, the DoS attack can be considered as a special case of the hybrid cyber-attacks, i.e., a DoS attack can either be equivalent to an FDI attack with false data being in the same magnitude of the actual signal but with an opposite sign or a jamming attack with high level noise variances such that the actual signal can be neglected compared to the noise signal \cite{Necip18}. On the other hand, if the jamming noise is correlated over the meters or it is not normally distributed, then such an attack does not comply with the considered jamming attack model in \eqref{eq:jamming_model} and nor with \eqref{eq:meas_attacked}. For such cases, we consider a non-parametric goodness-of-fit test as a countermeasure (see Sec.~\ref{subsec:counter}).

\subsection{Pre- and Post-Attack Measurement Densities}

Let $\mathbf{H} = [\mathbf{H}_1^\mathrm{T}, \mathbf{H}_2^\mathrm{T}, \dots, \mathbf{H}_K^\mathrm{T}]^\mathrm{T}$ where $\mathbf{H}_k \in \mathbb{R}^{\lambda \times N}$ is the measurement matrix for meter $k$. Since the measurement matrix is determined based on the system topology, the rows of $\mathbf{H}_k$ are identical, i.e., $\mathbf{H}_k = \mathbf{1}_{\lambda \times 1} \, \mathbf{h}_k^\mathrm{T}$, where $\mathbf{h}_k^\mathrm{T}$ is a row of $\mathbf{H}_k$. Based on the considered post-attack model in \eqref{eq:meas_attacked}, a measurement obtained at meter $k$ during the time interval between $t-1$ and $t$, i.e., $y_{k,t,i}, k \in \{1, 2, \dots, K\}, i \in \{1, 2, \dots, \lambda\}$ can be written as
\begin{equation}\label{eq:meas_att_v2}
y_{k,t,i} =
  \begin{cases}
    \mathbf{h}_k^\mathrm{T} \mathbf{x}_t + w_{k,t,i}, & \mbox{if } k \in \mathcal{S}_t^0 \\
    \mathbf{h}_k^\mathrm{T} \mathbf{x}_t + \mathrm{a}_{k,t} + w_{k,t,i}, & \mbox{if } k \in \mathcal{S}_t^f \\
    \mathbf{h}_k^\mathrm{T} \mathbf{x}_t + w_{k,t,i} + n_{k,t,i}, & \mbox{if } k \in \mathcal{S}_t^j \\
    \mathbf{h}_k^\mathrm{T} \mathbf{x}_t + \mathrm{a}_{k,t} + w_{k,t,i} + n_{k,t,i}, & \mbox{if } k \in \mathcal{S}_t^{f,j}
  \end{cases}
  , ~ t \geq \tau,
\end{equation}
where $\mathcal{S}_t^0$ is the set of non-attacked meters, $\mathcal{S}_t^f$ is the set of meters under only FDI attack, $\mathcal{S}_t^j$  is the set of meters under only jamming attack, and $\mathcal{S}_t^{f,j}$ is the set of meters under both FDI and jamming attacks at time $t \geq \tau$. Note that $\mathcal{S}_t^0$, $\mathcal{S}_t^f$, $\mathcal{S}_t^j$, and  $\mathcal{S}_t^{f,j}$ are disjoint sets and $\mathcal{S}_t^0 \cup \mathcal{S}_t^f \cup \mathcal{S}_t^j \cup \mathcal{S}_t^{f,j} = \{1,2,\dots,K\}$.

Then, the probability density functions (pdfs) of the measurements in the pre- and post-attack regimes take respectively the following forms $\forall i \in \{1,2,\dots,\lambda\}$:
\begin{align} \label{eq:hyp_null}
y_{k,t,i} \sim \mathbf{\mathcal{N}}(\mathbf{h}_k^\mathrm{T} \mathbf{x}_t,\sigma_w^2), ~~ \forall k \in \{1,2,\dots,K\}, ~ t < \tau,
\end{align}
and
\begin{align} \label{eq:hyp_alter}
y_{k,t,i} \sim
\begin{cases}
    \mathbf{\mathcal{N}}(\mathbf{h}_k^\mathrm{T} \mathbf{x}_t,\sigma_w^2), & \forall k \in \mathcal{S}_t^0 \\
    \mathbf{\mathcal{N}}(\mathbf{h}_k^\mathrm{T} \mathbf{x}_t + \mathrm{a}_{k,t},\sigma_w^2), & \forall k \in \mathcal{S}_t^f \\
    \mathbf{\mathcal{N}}(\mathbf{h}_k^\mathrm{T} \mathbf{x}_t,\sigma_w^2 + \sigma_{k,t}^2), & \forall k \in \mathcal{S}_t^j \\
    \mathbf{\mathcal{N}}(\mathbf{h}_k^\mathrm{T} \mathbf{x}_t + \mathrm{a}_{k,t},\sigma_w^2 + \sigma_{k,t}^2), & \forall k \in \mathcal{S}_t^{f,j}
  \end{cases}
  ,~ t \geq \tau.
\end{align}

\subsection{State Estimation}

Since the smart grid is modeled as a discrete-time linear dynamic system with the Gaussian noise terms (cf. \eqref{eq:state_upd} and \eqref{eq:meas_model}), the Kalman filter is the optimal linear estimator in minimizing the mean squared state estimation error \cite{Kalman_60}. Further, since the measurement models for the pre- and post-attack periods are different (cf. \eqref{eq:hyp_null} and \eqref{eq:hyp_alter}), two Kalman filters need to be simultaneously employed: one for assuming no attack occurs at all and one for  assuming an attack occurs at an unknown time $\tau$. Since the latter involves the unknown change-point $\tau$ and the unknown attack parameters $\mathbf{a}_t$ and $\pmb{\sigma}_t$, estimates of these unknowns are needed to employ the corresponding Kalman filter. As we will explain later, $\tau$ is estimated by the detection algorithm, $\mathbf{a}_t$ and $\pmb{\sigma}_t$ are estimated via the maximum likelihood (ML) estimation.

The Kalman filter is an iterative real-time estimator composed of prediction and measurement update steps at each iteration. Let the Kalman filter estimates for the pre- and post-attack cases be denoted with $\hat{\mathbf{x}}_{t|t'}^0$ and $\hat{\mathbf{x}}_{t|t'}^1$, respectively where $t' = t-1$ and $t' = t$ for the prediction and measurement update steps at time $t$, respectively. The Kalman filter equations at time $t$ are then given as follows:

\emph{Pre-attack -- Prediction}:
\begin{gather} \nonumber
\hat{\mathbf{x}}_{t|t-1}^0 = \mathbf{A} \hat{\mathbf{x}}_{t-1|t-1}^0, \\ \label{eq:pred_fdata_null}
\mathbf{P}_{t|t-1}^0 = \mathbf{A} \mathbf{P}_{t-1|t-1}^0 \mathbf{A}^\mathrm{T} + \sigma_v^2 \, \mathbf{I}_N,
\end{gather}

\emph{Pre-attack -- Measurement update}:
\begin{gather} \nonumber
\mathbf{G}_{t}^0 = \mathbf{P}_{t|t-1}^0 \mathbf{H}^\mathrm{T} (\mathbf{H} \mathbf{P}_{t|t-1}^0 \mathbf{H}^\mathrm{T} + \sigma_w^2 \, \mathbf{I}_{K \lambda})^{-1}, \\ \nonumber
\hat{\mathbf{x}}_{t|t}^0 = \hat{\mathbf{x}}_{t|t-1}^0 + \mathbf{G}_{t}^0 (\mathbf{y}_t - \mathbf{H} \hat{\mathbf{x}}_{t|t-1}^0), \\ \label{eq:meas_upd_fdata_null}
\mathbf{P}_{t|t}^0 = \mathbf{P}_{t|t-1}^0 - \mathbf{G}_{t}^0 \mathbf{H} \mathbf{P}_{t|t-1}^0,
\end{gather}

\emph{Post-attack -- Prediction}:
\begin{gather} \nonumber
\hat{\mathbf{x}}_{t|t-1}^1 = \mathbf{A} \hat{\mathbf{x}}_{t-1|t-1}^1, \\ \label{eq:pred_fdata_alter}
\mathbf{P}_{t|t-1}^1 = \mathbf{A} \mathbf{P}_{t-1|t-1}^1 \mathbf{A}^\mathrm{T} + \sigma_v^2 \, \mathbf{I}_N,
\end{gather}

\emph{Post-attack -- Measurement update}:
\begin{gather} \nonumber
\mathbf{G}_{t}^1 = \mathbf{P}_{t|t-1}^1 \mathbf{H}^\mathrm{T} (\mathbf{H} \mathbf{P}_{t|t-1}^1 \mathbf{H}^\mathrm{T} + \sigma_w^2 \, \mathbf{I}_{K \lambda} + \mathrm{diag}(\hat{\pmb{\sigma}}_t))^{-1}, \\ \nonumber
\hat{\mathbf{x}}_{t|t}^1 = \hat{\mathbf{x}}_{t|t-1}^1 + \mathbf{G}_{t}^1 (\mathbf{y}_t - \mathbf{H} \hat{\mathbf{x}}_{t|t-1}^1 - \hat{\mathbf{a}}_t), \\ \label{eq:meas_upd_fdata_alter}
\mathbf{P}_{t|t}^1 = \mathbf{P}_{t|t-1}^1 - \mathbf{G}_{t}^1 \mathbf{H} \mathbf{P}_{t|t-1}^1,
\end{gather}
where $\mathbf{P}_{t|t'}^0$ and $\mathbf{P}_{t|t'}^1$ denote the estimates of the state covariance matrix at time $t$, and $\mathbf{G}_{t}^0$ and $\mathbf{G}_{t}^1$ denote the Kalman gain matrices at time $t$ for the pre- and post-attack cases, respectively. Note that the MLE estimates of the attack parameters are used in the measurement update step of the Kalman filter for the post-attack case, where $\hat{\mathbf{a}}_t$ is the MLE of $\mathbf{a}_t$ (cf. \eqref{eq:a_hat_kt_v3_rrr}) and $\hat{\pmb{\sigma}}_t$ is the MLE of $\pmb{\sigma}_t$ (cf. \eqref{eq:sigma_hat_kt_v3_rrr}). Hence, $\hat{\mathbf{x}}_{t|t-1}^1$ and $\hat{\mathbf{x}}_{t|t}^1$ are, in fact, recovered state estimates in case of a cyber-attack. Note, however, that ML estimation errors may lead to errors in computing the recovered state estimates.

\subsection{Problem Formulation}

Our objective is detecting cyber-attacks in a timely and reliable manner and the quickest detection theory \cite{Poor08,Basseville93,Veeravalli14} is well suited to this objective. In the quickest change detection problems, measurements become available sequentially over time and at each time, either a change is declared or further measurements are taken in the next time interval, where the aim is to optimally balance the detection delay and the false alarm rate. There are two main approaches in the quickest detection theory, namely Bayesian and non-Bayesian. In a Bayesian setting, the change point $\tau$ is considered as a random variable with a known a priori distribution whereas in a non-Bayesian setting, the change point is considered as non-random and unknown. Our problem better fits to the non-Bayesian setting since we do not assume any a priori knowledge about the change-point $\tau$. Then, we consider the following objective function, proposed by Lorden \cite{Lorden_71}:
\begin{gather} \label{eq:det_delay}
d(T) = \sup_{\tau} \, \esssup_{\mathcal{F}_\tau} \, \mathbb{E}_\tau \big[(T-\tau)^+\,|\mathcal{F}_\tau\,\big],
\end{gather}
where $T$ is the stopping time at which an attack is declared, $\mathcal{F}_\tau$ denotes all measurements obtained up to time $\tau$, and $\mathbb{E}_j$ is the expectation under $\mathbb{P}_j$, that is the probability measure if the change occurs at time $j$. Note that $d(T)$ is called the worst-case average detection delay since it is maximized over the change point and also over all measurements obtained up to the change-point. We then consider the following minimax optimization problem:
\begin{gather} \label{eq:opt_prob}
\inf_{T}~ d(T) ~~ \text{subject to} ~~ \mathbb{E}_\infty[T] \geq \alpha,
\end{gather}
where $\mathbb{E}_\infty[T]$ is called the average false alarm period, i.e., average stopping time when no change occurs at all ($\tau = \infty$), and $\alpha$ is a prespecified lower bound for $\mathbb{E}_\infty[T]$.

Let the pre- and post-attack measurement pdfs given in \eqref{eq:hyp_null} and \eqref{eq:hyp_alter} be denoted with $p_0(\mathbf{y}_t|\mathbf{x}_t)$ and $p_1(\mathbf{y}_t|\mathbf{x}_t, \mathbf{a}_t, \pmb{\sigma}_t)$, respectively. Since the dynamic system state $\mathbf{x}_t$ is not directly observed and the attack parameters $\mathbf{a}_t$ and $\pmb{\sigma}_t$ are completely determined by an attacker and hence unknown, both pdfs are unknown and time-varying. If the pre- and post-attack pdfs would be exactly known, then the well-known CUSUM algorithm would be the optimal solution to \eqref{eq:opt_prob} \cite{Moustakides_86}. Nonetheless, the system state can be inferred using the Kalman filters and the MLEs of the unknown attack parameters can be computed. Then, following a generalized likelihood ratio approach \cite[Sec.~5.3]{Basseville93}, \cite{Li_15,Necip18} and replacing the unknowns with their estimates, a generalized CUSUM algorithm can be used as a solution to \eqref{eq:opt_prob}.

In this paper, in addition to early attack detection, we also aim to recover the attack-free system states. Notice that in case of no attack, i.e., for $t < \tau$, the Kalman filter for the pre-attack case (assuming no attack at all) is the optimal state estimator. However, after an attack occurs, the measurement model assumed in the pre-attack period (cf. \eqref{eq:meas_model}) is no longer true. Hence, the state estimates for the pre-attack case, i.e., $\hat{\mathbf{x}}_{t|t-1}^0$ and $\hat{\mathbf{x}}_{t|t}^0$, deviate from the actual system state $\mathbf{x}_t$ for $t \geq \tau$. Recalling that an attack occurs at an unknown time $\tau$ and the measurements follow the post-attack measurement model (cf. \eqref{eq:meas_attacked}) for $t \geq \tau$, if the attack launch time $\tau$ and the attack magnitudes $\mathbf{a}_t$ and $\pmb{\sigma}_t$ would be exactly known, then the system state would be perfectly recovered for $t \geq \tau$. Nonetheless, as we will explain more clearly in the next section, the (generalized) CUSUM algorithm always keeps a change-point estimate $\hat{\tau}$ in its memory and updates this estimate as the measurements become sequentially available over time \cite[Sec. 2.2]{Basseville93}. When an attack is declared at the stopping time $T$, $\hat{\tau}$ becomes the final change-point estimate of the (generalized) CUSUM algorithm. Furthermore, the MLEs of the attack magnitudes, i.e., $\hat{\mathbf{a}}_t$ and $\hat{\pmb{\sigma}}_t$, can be computed at each time $t$. Then, employing a Kalman filter for the post-attack case (cf. \eqref{eq:pred_fdata_alter} and \eqref{eq:meas_upd_fdata_alter}) and computing the state estimates using the MLEs of the attack parameters in the measurement update step, recovered state estimates, i.e., $\hat{\mathbf{x}}_{t|t-1}^1$ and $\hat{\mathbf{x}}_{t|t}^1$, can be obtained for $\hat{\tau} \leq t \leq T$.

\section{Online Attack Detection and Estimation} \label{sec:combined}

Since it is hard to distinguish noise from FDI/jamming attacks with small magnitudes, some minimum levels for the attack magnitudes need to be defined in order to control the false alarm level of a detection algorithm. We then define the change event of interest as follows:
\begin{gather} \nonumber
 |\mathrm{a}_{k,t}| \geq \gamma,~ \forall k \in \mathcal{S}_t^f, ~~ t \geq \tau, \\ \nonumber
 \sigma_{k,t}^2 \geq \sigma^2,~ \forall k \in \mathcal{S}_t^j, ~~ t \geq \tau, \\ \label{eq:change_event}
 |\mathrm{a}_{k,t}| \geq \gamma ~ \& ~ \sigma_{k,t}^2 \geq \sigma^2, ~ \forall k \in \mathcal{S}_t^{f,j}, ~~ t \geq \tau,
\end{gather}
where $\gamma$ and $\sigma^2$ are the smallest attack magnitudes of interest for $|\mathrm{a}_{k,t}|$ and $\sigma_{k,t}^2$, respectively. Note that, in general, an attacker can arbitrarily choose its attack parameters, i.e., $\gamma$ and $\sigma^2$ do not restrict an attacker's strategy. In fact, attackers usually do not know such parameters. On the other hand, smarter attackers may exploit such lower bounds on the attack magnitudes in order to perform stealthy attacks with small attack magnitudes (see Sec.~\ref{sec:st_proposed}).

The generalized CUSUM algorithm can then be written as follows:
\begin{gather} \label{eq:gen_CUSUM}
T = \inf \bigg\{m \in \mathbb{N}: \underbrace{ \max_{1 \leq j \leq m} \sum_{t=j}^{m} \underbrace{ \sup_{\mathcal{S}_t^0,\mathcal{S}_t^f,\mathcal{S}_t^j,\mathcal{S}_t^{f,j}} \, \log {\frac{\sup_{|a_{k,t}| \geq \gamma, \, k \, \in \, \mathcal{S}_t^f \, \cup \, \mathcal{S}_t^{f,j}} \, \sup_{\sigma_{k,t}^2 \geq \sigma^2, \, k \, \in \, \mathcal{S}_t^j \, \cup \, \mathcal{S}_t^{f,j}} \, p_1(\mathbf{y}_t \,|\, \hat{\mathbf{x}}_t^1, \mathbf{a}_t, \pmb{\sigma}_t) }{p_0(\mathbf{y}_t|\hat{\mathbf{x}}_t^0)}}}_{\beta_t}}_{g_m} \geq h \bigg\},
\end{gather}
where $\hat{\mathbf{x}}_t^0$ and $\hat{\mathbf{x}}_t^1$ denote the state estimates for the pre- and post-attack cases, respectively, $g_m$ is the decision statistic at time $m$, $h$ is the test threshold, and $\beta_t$ is the generalized log-likelihood ratio (GLLR) calculated at time $t$. Based on \eqref{eq:gen_CUSUM}, the decision statistic can be recursively updated at each time $t$ as $g_t \gets \max\{0, g_{t-1} + \beta_t\}$, where $g_0 = 0$ \cite[Sec. 2.2]{Basseville93}.

Note that whenever $g_t$ reaches zero, the (generalized) CUSUM algorithm updates its change-point estimate $\hat{\tau}$ to the current time $t$, where the initial change-point estimate is $\hat{\tau} = 1$ \cite[Sec. 2.2]{Basseville93}. That is, when $g_t \gets 0$, we have $\hat{\tau} \gets t$. Recall that the Kalman filter for the post-attack case is employed assuming the normal measurement model (cf. \eqref{eq:meas_model}) up to the unknown change-point $\tau$. We then propose to employ the Kalman filter for the post-attack case based on the estimated change-point $\hat{\tau}$. Hence, whenever the change-point estimate is updated, the Kalman filter for the post-attack case needs also to be updated. Recall further that the Kalman filter for the pre-attack case is always employed based on the normal measurement model. Hence, whenever $g_t \gets 0$, the Kalman filter estimates for the post-attack case are updated by setting them to the Kalman filter estimates for the pre-attack case, i.e., $\mathbf{P}_{t|t}^1 \gets \mathbf{P}_{t|t}^0$ and $\hat{\mathbf{x}}_{t|t}^1 \gets \hat{\mathbf{x}}_{t|t}^0$.

Assuming no attack, $\hat{\mathbf{x}}_{t|t}^0$ is the optimal state estimate at time $t$. Thus, we estimate $\mathbf{x}_t$ by $\hat{\mathbf{x}}_{t|t}^0$ for the pre-attack case, i.e., $\hat{\mathbf{x}}_t^0 \triangleq \hat{\mathbf{x}}_{t|t}^0$. On the other hand, we estimate $\mathbf{x}_t$ by $\hat{\mathbf{x}}_{t|t-1}^1$ for the post-attack case, i.e., $\hat{\mathbf{x}}_t^1 \triangleq \hat{\mathbf{x}}_{t|t-1}^1$. This is because the measurement update step of the Kalman filter for the post-attack case and hence $\hat{\mathbf{x}}_{t|t}^1$ depends on estimates of the unknown attack variables (cf. \eqref{eq:meas_upd_fdata_alter}), and effects of the attack parameters $\mathbf{a}_t$ and $\pmb{\sigma}_t$ at time $t$ on $\hat{\mathbf{x}}_t^1$ need to be blocked to be able to compute the MLEs of the attack parameters in closed form (cf. numerator in \eqref{eq:gen_CUSUM}).  Note that $\hat{\mathbf{x}}_{t|t-1}^1$ is computed based on the measurements up to time $t-1$, thus $\hat{\mathbf{x}}_{t|t-1}^1$ is independent of the attack parameters at time $t$.

At first, it may seem unfair that we use the state estimate of the measurement update step, i.e., $\hat{\mathbf{x}}_{t|t}^0$, for the pre-attack case, and the state prediction, i.e., $\hat{\mathbf{x}}_{t|t-1}^1$, for the post-attack case. However, it essentially improves the performance of the proposed detection scheme due to the following reasons: (i) in case of no attack, we favor $p_0(\mathbf{y}_t|\hat{\mathbf{x}}_t^0)$ over $p_1(\mathbf{y}_t \,|\, \hat{\mathbf{x}}_t^1, \mathbf{a}_t, \pmb{\sigma}_t)$ and hence decrease the false alarm level of the proposed detection scheme, (ii) in case of an attack, since the state estimates for the post-attack case are recovered whereas the state estimates for the pre-attack case do not have a recovery mechanism, detection delays are not expected to increase.


Furthermore, based on \eqref{eq:gen_CUSUM}, the following proposition presents the GLLR at time $t$ and the MLEs of the attack variables for the time interval between $t-1$ and $t$.

\textbf{Proposition 1:} Let $e_{k,t,i} \triangleq y_{k,t,i} - \mathbf{h}_k^\mathrm{T} \hat{\mathbf{x}}_t^1$ and $\mathbf{e}_{k,t} \triangleq [e_{k,t,1}, e_{k,t,2}, \dots, e_{k,t,\lambda}]^\mathrm{T}$. Moreover, let $\delta_{k,t} \triangleq \sum_{i=1}^{\lambda} e_{k,t,i}$, $\zeta_{k,t} \triangleq \sum_{i=1}^{\lambda} e_{k,t,i}^2$, $\varrho_{k,t} \triangleq \sum_{i=1}^{\lambda} (e_{k,t,i} + \gamma)^2$, and $\varpi_{k,t} \triangleq \sum_{i=1}^{\lambda} (e_{k,t,i} - \gamma)^2$, $\forall k \in \{1,2,\dots,K\}$, $\forall t > 0$. The most likely subset of meters under no attack, under only FDI attack, under only jamming attack, and under both FDI and jamming attacks during the time interval between $t-1$ and $t$ are classified, respectively as
\begin{align} \label{eq:S0_rrr}
&\hat{\mathcal{S}}_t^0 = \Big\{k: u^0(\mathbf{e}_{k,t}) \leq u^f(\mathbf{e}_{k,t}), \, u^0(\mathbf{e}_{k,t}) \leq u^j(\mathbf{e}_{k,t}), \, u^0(\mathbf{e}_{k,t}) \leq  u^{f,j}(\mathbf{e}_{k,t}), \, k = 1, 2, \dots, K \Big\}, \\ \label{eq:Sf_rrr}
&\hat{\mathcal{S}}_t^f = \Big\{ k: u^f(\mathbf{e}_{k,t}) < u^0(\mathbf{e}_{k,t}), \, u^f(\mathbf{e}_{k,t}) \leq u^j(\mathbf{e}_{k,t}), \, u^f(\mathbf{e}_{k,t}) \leq u^{f,j}(\mathbf{e}_{k,t}), \, k = 1, 2, \dots, K \Big\}, \\ \label{eq:Sj_rrr}
&\hat{\mathcal{S}}_t^j = \Big\{ k: u^j(\mathbf{e}_{k,t}) < u^0(\mathbf{e}_{k,t}), \, u^j(\mathbf{e}_{k,t}) < u^f(\mathbf{e}_{k,t}), \, u^j(\mathbf{e}_{k,t}) \leq u^{f,j}(\mathbf{e}_{k,t}), \, k = 1, 2, \dots, K \Big\}, \\ \label{eq:Sfj_rrr}
&\hat{\mathcal{S}}_t^{f,j} = \Big\{ k: u^{f,j}(\mathbf{e}_{k,t}) < u^0(\mathbf{e}_{k,t}), \, u^{f,j}(\mathbf{e}_{k,t}) < u^f(\mathbf{e}_{k,t}), \, u^{f,j}(\mathbf{e}_{k,t}) < u^j(\mathbf{e}_{k,t}), \, k = 1, 2, \dots, K \Big\},
\end{align}
and the GLLR at time $t$ is computed as
\begin{align} \nonumber
\beta_t &=  \frac{K \lambda}{2} \log(\sigma_w^2) + \frac{1}{2 \sigma_w^2} \sum_{k=1}^{K} \sum_{i=1}^{\lambda} (y_{k,t,i} - \mathbf{h}_k^\mathrm{T} \hat{\mathbf{x}}_t^0)^2 \\ \label{eq:beta_t_v3_rrr}
&~~ - \frac{1}{2} \bigg( \sum_{k \in \hat{\mathcal{S}}_t^0} u^0(\mathbf{e}_{k,t}) + \sum_{k \in \hat{\mathcal{S}}_t^f} u^f(\mathbf{e}_{k,t})
+ \sum_{k \in \hat{\mathcal{S}}_t^j} u^j(\mathbf{e}_{k,t}) + \sum_{k \in \hat{\mathcal{S}}_t^{f,j}} u^{f,j}(\mathbf{e}_{k,t}) \bigg),
\end{align}
where
\begin{gather}\label{eq:u0_rrr}
u^0(\mathbf{e}_{k,t}) \triangleq  \lambda \log(\sigma_w^2) + \frac{\zeta_{k,t}}{\sigma_w^2},
\end{gather}
\begin{align} \label{eq:uf_rrr}
u^f(\mathbf{e}_{k,t}) &\triangleq \begin{cases}
  \lambda \log(\sigma_w^2) + \frac{1}{\sigma_w^2} \sum_{i=1}^{\lambda} (e_{k,t,i} - \frac{\delta_{k,t}}{\lambda})^2 , & \mbox{if } |\frac{\delta_{k,t}}{\lambda}| \geq \gamma \\
  \lambda \log(\sigma_w^2) + \frac{\varpi_{k,t}}{\sigma_w^2}, & \mbox{if } 0 \leq \frac{\delta_{k,t}}{\lambda} < \gamma \\
  \lambda \log(\sigma_w^2) + \frac{\varrho_{k,t}}{\sigma_w^2}, & \mbox{if } -\gamma < \frac{\delta_{k,t}}{\lambda} < 0,
\end{cases}
\end{align}
\begin{align} \label{eq:uj_rrr}
u^j(\mathbf{e}_{k,t}) &\triangleq \begin{cases}
  \lambda \log(\frac{\zeta_{k,t}}{\lambda}) + \lambda, & \mbox{if } \frac{\zeta_{k,t}}{\lambda} \geq \sigma_w^2 + \sigma^2 \\
  \lambda \log(\sigma_w^2 + \sigma^2) + \frac{\zeta_{k,t}}{\sigma_w^2 + \sigma^2}, & \mbox{if } \frac{\zeta_{k,t}}{\lambda} < \sigma_w^2 + \sigma^2,
\end{cases}
\end{align}
and
\begin{align} \label{eq:ufj_rrr}
u^{f,j}(\mathbf{e}_{k,t}) &\triangleq
 \begin{cases}
     \lambda \log(\frac{1}{\lambda} \sum_{i=1}^{\lambda} (e_{k,t,i} - \frac{\delta_{k,t}}{\lambda})^2) + \lambda, & \mbox{if } |\frac{\delta_{k,t}}{\lambda}| \geq \gamma \mbox{ and } \frac{1}{\lambda} \sum_{i=1}^{\lambda} (e_{k,t,i} - \frac{\delta_{k,t}}{\lambda})^2 \geq \sigma_w^2 + \sigma^2 \\
     \lambda \log(\sigma_w^2 + \sigma^2) + \frac{1}{\sigma_w^2 + \sigma^2} \sum_{i=1}^{\lambda} (e_{k,t,i} - \frac{\delta_{k,t}}{\lambda})^2, & \mbox{if } |\frac{\delta_{k,t}}{\lambda}| \geq \gamma \mbox{ and } \frac{1}{\lambda} \sum_{i=1}^{\lambda} (e_{k,t,i} - \frac{\delta_{k,t}}{\lambda})^2 < \sigma_w^2 + \sigma^2 \\
     \lambda \log(\frac{\varpi_{k,t}}{\lambda}) + \lambda, & \mbox{if } 0 \leq \frac{\delta_{k,t}}{\lambda} < \gamma \mbox{ and } \frac{\varpi_{k,t}}{\lambda} \geq \sigma_w^2 + \sigma^2 \\
     \lambda \log(\sigma_w^2 + \sigma^2) + \frac{\varpi_{k,t}}{\sigma_w^2 + \sigma^2}, & \mbox{if } 0 \leq \frac{\delta_{k,t}}{\lambda} < \gamma \mbox{ and } \frac{\varpi_{k,t}}{\lambda} < \sigma_w^2 + \sigma^2 \\
     \lambda \log(\frac{\varrho_{k,t}}{\lambda}) + \lambda, & \mbox{if } -\gamma < \frac{\delta_{k,t}}{\lambda} < 0 \mbox{ and } \frac{\varrho_{k,t}}{\lambda} \geq \sigma_w^2 + \sigma^2 \\
     \lambda \log(\sigma_w^2 + \sigma^2) + \frac{\varrho_{k,t}}{\sigma_w^2 + \sigma^2}, & \mbox{if } -\gamma < \frac{\delta_{k,t}}{\lambda} < 0 \mbox{ and } \frac{\varrho_{k,t}}{\lambda} < \sigma_w^2 + \sigma^2.
   \end{cases}
\end{align}
Furthermore, the MLEs of the attack magnitudes for meter $k \in \{1,2,\dots,K\}$ and for the interval between $t-1$ and $t$ are determined as follows:
\begin{equation} \label{eq:a_hat_kt_v3_rrr}
    \hat{\mathrm{a}}_{k,t} =
    \begin{cases}
     \frac{\delta_{k,t}}{\lambda}, & \text{if} ~~ |\frac{\delta_{k,t}}{\lambda}| \geq \gamma \mbox{ and } k \in \hat{\mathcal{S}}_t^f \cup \hat{\mathcal{S}}_t^{f,j} \\
     \gamma , & \text{if} ~~ 0 \leq \frac{\delta_{k,t}}{\lambda} < \gamma \mbox{ and } k \in \hat{\mathcal{S}}_t^f \cup \hat{\mathcal{S}}_t^{f,j} \\
     - \gamma , & \text{if} ~~ -\gamma < \frac{\delta_{k,t}}{\lambda} < 0 \mbox{ and } k \in \hat{\mathcal{S}}_t^f \cup \hat{\mathcal{S}}_t^{f,j} \\
     0, & \text{if} ~~ k \in \hat{\mathcal{S}}_t^0 \cup \hat{\mathcal{S}}_t^j
    \end{cases}
\end{equation}
and
\begin{align} \label{eq:sigma_hat_kt_v3_rrr}
\hat{\sigma}_{k,t}^2 &=
    \begin{cases}
     - \sigma_w^2 + \frac{\zeta_{k,t}}{\lambda}, & \text{if} ~~ \frac{\zeta_{k,t}}{\lambda} \geq \sigma_w^2 + \sigma^2 \mbox{ and } k \in \hat{\mathcal{S}}_t^j  \\
     \sigma^2 , & \text{if} ~~ \frac{\zeta_{k,t}}{\lambda} < \sigma_w^2 + \sigma^2 \mbox{ and } k \in \hat{\mathcal{S}}_t^j \\
     - \sigma_w^2 + \frac{1}{\lambda} \sum_{i=1}^{\lambda} (e_{k,t,i} - \frac{\delta_{k,t}}{\lambda})^2, & \mbox{if } |\frac{\delta_{k,t}}{\lambda}| \geq \gamma \mbox{ and } \frac{1}{\lambda} \sum_{i=1}^{\lambda} (e_{k,t,i} - \frac{\delta_{k,t}}{\lambda})^2 \geq \sigma_w^2 + \sigma^2 \mbox{ and } k \in \hat{\mathcal{S}}_t^{f,j} \\
     \sigma^2, & \mbox{if } |\frac{\delta_{k,t}}{\lambda}| \geq \gamma \mbox{ and } \frac{1}{\lambda} \sum_{i=1}^{\lambda} (e_{k,t,i} - \frac{\delta_{k,t}}{\lambda})^2 < \sigma_w^2 + \sigma^2 \mbox{ and } k \in \hat{\mathcal{S}}_t^{f,j} \\
     - \sigma_w^2 + \frac{\varpi_{k,t}}{\lambda}, & \mbox{if } 0 \leq \frac{\delta_{k,t}}{\lambda} < \gamma \mbox{ and } \frac{\varpi_{k,t}}{\lambda} \geq \sigma_w^2 + \sigma^2 \mbox{ and } k \in \hat{\mathcal{S}}_t^{f,j} \\
     \sigma^2, & \mbox{if } 0 \leq \frac{\delta_{k,t}}{\lambda} < \gamma \mbox{ and } \frac{\varpi_{k,t}}{\lambda} < \sigma_w^2 + \sigma^2 \mbox{ and } k \in \hat{\mathcal{S}}_t^{f,j} \\
     - \sigma_w^2 + \frac{\varrho_{k,t}}{\lambda}, & \mbox{if } -\gamma < \frac{\delta_{k,t}}{\lambda} < 0 \mbox{ and } \frac{\varrho_{k,t}}{\lambda} \geq \sigma_w^2 + \sigma^2 \mbox{ and } k \in \hat{\mathcal{S}}_t^{f,j} \\
     \sigma^2, & \mbox{if } -\gamma < \frac{\delta_{k,t}}{\lambda} < 0 \mbox{ and } \frac{\varrho_{k,t}}{\lambda} < \sigma_w^2 + \sigma^2 \mbox{ and } k \in \hat{\mathcal{S}}_t^{f,j} \\
     0, & \mbox{if } k \in \hat{\mathcal{S}}_t^0 \cup \hat{\mathcal{S}}_t^f.
    \end{cases}
\end{align}

\textbf{Proof:} See Appendix~\ref{sec:proof_prop1}.

The proposed online detection and estimation algorithm is summarized in Algorithm~\ref{alg:centralized}. At each time $t$, firstly the prediction step of the Kalman filters is implemented. Then, the most likely attack type (or no attack) and the attack parameters for each meter are determined. Based on the estimates of the attack variables, the measurement update step of the Kalman filters is implemented. Then, the GLLR is computed and the decision statistic is updated. If the decision statistic crosses the predetermined test threshold, then an attack is declared. Otherwise, it proceeds to the next time interval and further measurements are collected. Moreover, if the decision statistic reaches zero, the Kalman filter estimates for the post-attack case are updated before proceeding to the next time interval. Recall that Algorithm 1 keeps a change point estimate $\hat{\tau}$. Hence, after an attack is declared at time $T$, to help for a quick system recovery, $\{\hat{\mathbf{x}}_{t|t}^1: \hat{\tau} \leq t \leq T\}$ can be reported as the recovered state estimates and further, estimates of the attack types and the set of attacked meters can be reported for the time interval between $\hat{\tau}$ and $T$.

\begin{algorithm}[t]\small
\caption{\small Real-time attack detection and estimation}
\label{alg:centralized}
\baselineskip=0.4cm
\begin{algorithmic}[1]
\STATE Initialization: $t \gets 0$, $g_{0} \gets 0$, $\hat{\tau} \gets 1$
\WHILE {$g_t < h$}
    \STATE $t \gets t+1$
    \STATE Implement the prediction step of the Kalman filters using \eqref{eq:pred_fdata_null} and \eqref{eq:pred_fdata_alter}.
    \STATE Compute $u^0(\mathbf{e}_{k,t}), u^f(\mathbf{e}_{k,t}), u^j(\mathbf{e}_{k,t}), \mbox{ and } u^{f,j}(\mathbf{e}_{k,t}), \, \forall k \in \{1, 2, \dots, K\}$ using \eqref{eq:u0_rrr}, \eqref{eq:uf_rrr}, \eqref{eq:uj_rrr}, and \eqref{eq:ufj_rrr}, respectively.
    \STATE Classification: compute $\hat{\mathcal{S}}_t^0, \hat{\mathcal{S}}_t^f, \hat{\mathcal{S}}_t^j, \mbox{ and } \hat{\mathcal{S}}_t^{f,j}$ using \eqref{eq:S0_rrr}, \eqref{eq:Sf_rrr}, \eqref{eq:Sj_rrr}, and \eqref{eq:Sfj_rrr}, respectively.
    \STATE Compute $\hat{\mathbf{a}}_t$ and $\hat{\pmb{\sigma}}_t$ using \eqref{eq:a_hat_kt_v3_rrr} and \eqref{eq:sigma_hat_kt_v3_rrr}, respectively.
    \STATE Implement the measurement update step of the Kalman filters using \eqref{eq:meas_upd_fdata_null} and \eqref{eq:meas_upd_fdata_alter}.
    \STATE Compute $\beta_t$ using \eqref{eq:beta_t_v3_rrr}.
    \STATE Update the decision statistic: $g_t \gets \max\{0, g_{t-1} + \beta_t\}$
    \IF {$g_t = 0$}
        \STATE $\hat{\tau} \gets t$
        \STATE $\hat{\mathbf{x}}_{t|t}^1 \gets \hat{\mathbf{x}}_{t|t}^0$
        \STATE $\mathbf{P}_{t|t}^1 \gets \mathbf{P}_{t|t}^0$
	\ENDIF
\ENDWHILE
\STATE $T \gets t$, declare a cyber-attack.
\end{algorithmic}
\end{algorithm}

\textit{Remark 2:} The detector parameters $\gamma$ and $\sigma^2$ can be determined by the system designer based on the system requirements, i.e., the desired level of false alarm rate. The system designer firstly determines the desired minimum level of average false alarm period, i.e., $\alpha$. If the frequency of false alarms needs to be decreased, then $\alpha$ is chosen higher. After choosing $\alpha$, the system designer chooses the values of $\gamma$, $\sigma^2$, and the test threshold $h$ in order to achieve an average false alarm period that is larger than or equal to $\alpha$. For a higher level of $\alpha$, higher values of $\gamma$, $\sigma^2$, and $h$ need to be chosen. On the other hand, higher values of $\gamma$, $\sigma^2$, and $h$ lead to larger detection delays. Hence, the system designer can choose such parameters to strike a desired balance between false alarm rate and the detection delays.

\section{Stealthy Attacks and Countermeasures} \label{sec:stealth}

We firstly discuss stealthy attacks against a CUSUM detector, which can be employed in a simple case where the pre- and post-attack pdfs are known. Discussion on the stealthy attacks against a CUSUM detector is useful since similar forms of stealthy attacks can be performed against all CUSUM-based detectors. We then particularly discuss stealthy attacks against the proposed detector, i.e., Algorithm 1, where the pre- and post-attack pdfs are unknown and time-varying, as explained in Sec.~\ref{sec:combined}. Finally, we present some countermeasures against the considered stealthy attacks.

\subsection{Stealthy Attacks Against a CUSUM Detector} \label{sec:st_cusum}

Suppose the pre- and post-attack measurement pdfs are known and denoted with $f_0$ and $f_1$, respectively such that $y_t \sim f_0$ for $t < \tau$ and $y_t \sim f_1$ for $t \geq \tau$. In this case, the CUSUM algorithm is the optimum solution to \eqref{eq:opt_prob} \cite{Moustakides_86}, given by
\begin{gather}\label{eq:cusum}
T_{\text{CUSUM}} = \inf\{t: g_t \geq h\}, ~ g_t = \max\{0, g_{t-1} + \ell_t\},
\end{gather}
where $T_{\text{CUSUM}}$ denotes the stopping time, $h$ is the test threshold, $g_t$ is the decision statistic at time $t$, and $\ell_t \triangleq \log\big(\frac{f_1(y_t)}{f_0(y_t)}\big)$ is the log-likelihood ratio (LLR) at time $t$.


\noindent \subsubsection{Non-persistent attacks}

The CUSUM algorithm is mainly designed for detecting persistent changes, i.e., it is assumed that an attack is launched at an unknown time $\tau$ and continued thereafter. It accumulates evidence (LLR) over time and declares a change (attack/anomaly) only if the accumulated evidence is reliably high (cf. \eqref{eq:cusum}). Hence, with the purpose of increasing the detection delay of the CUSUM algorithm, a smart attacker can design an on-off attacking strategy to perform an intermittent (non-persistent) attack.  That is, it can attack for a period of time, then wait for a period of time and repeat this procedure over its attacking period with the aim of keeping the decision statistic of the CUSUM algorithm, i.e., $g_t$, below the decision threshold $h$ for $t \geq \tau$ so that the attack can be continued without being noticed.

Since the measurements $y_t$ are essentially random variables, an attacker cannot control the decision statistic deterministically; it can control it only on average. Note that attackers usually need simple and effective attacking strategies that require the minimum possible knowledge. Let $\mathrm{KL}(f_1,f_0) \triangleq \int{f_1(y) \log(\frac{f_1(y)}{f_0(y)}) dy}$ denote the Kullback-Leibler (KL) divergence between $f_1$ and $f_0$. The following proposition presents a simple necessary condition for an attacker, having the knowledge of $f_0$ and $f_1$, to determine the on and off periods of a non-persistent stealthy attack against the CUSUM detector.

\textbf{Proposition 2:} Let $h' \geq \mathrm{KL}(f_1,f_0)$ be a threshold chosen by the attacker. The on and off periods have to be chosen as
\begin{gather}\nonumber
\mathrm{T}_{\text{on}} \leq \frac{h'}{\mathrm{KL}(f_1,f_0)} \mbox{ and } \mathrm{T}_{\text{off}} > \frac{h'}{\mathrm{KL}(f_0,f_1)}
\end{gather}
in order to satisfy $\mathbb{E}[g_t] \leq h'$ for $t \geq \tau$, where $\mathrm{T}_{\text{on}}$ and $\mathrm{T}_{\text{off}}$ are positive integers denoting the on and off periods, respectively.

\begin{proof}

We have
\begin{align}\nonumber
 \mathbb{E}[g_t] &= \mathbb{E}[\max\{0, g_{t-1} + \ell_t\}] \\ \label{eq:low_bnd}
  &\geq  \max\{0, \mathbb{E}[g_{t-1} + \ell_t]\} = \max\{0, \mathbb{E}[g_{t-1}] + \mathbb{E}[\ell_t]\},
\end{align}
where the inequality is due to the fact that $g_{t-1} + \ell_t$ can take negative values in general ($-\infty < \ell_t < \infty$).

If $y_t \sim f_1$, then
\begin{gather} \nonumber
\mathbb{E}[\ell_t] = \int{f_1(y) \log(\frac{f_1(y)}{f_0(y)}) dy} = \mathrm{KL}(f_1,f_0) > 0,
\end{gather}
and if $y_t \sim f_0$, then
\begin{gather} \nonumber
\mathbb{E}[\ell_t] = \int{f_0(y) \log(\frac{f_1(y)}{f_0(y)}) dy} = -\mathrm{KL}(f_0,f_1) < 0.
\end{gather}
Let
\begin{align} \label{eq:rho_t}
\rho_t \triangleq \max\{0, \mathbb{E}[g_{t-1}] + \mathbb{E}[\ell_t]\}
\end{align}
be a lower bound on $\mathbb{E}[g_t]$ (cf. \eqref{eq:low_bnd}). Since (i) $g_t = 0$ at $t = 0$ (hence, $\mathbb{E}[g_0] = 0$) and (ii) for $t \leq \tau-1$, $\mathbb{E}[\ell_t] = -\mathrm{KL}(f_0,f_1) < 0$, based on \eqref{eq:rho_t}, we have $\rho_t = 0$ for $t \leq \tau-1$. Further, based on \eqref{eq:rho_t}, with an on period of $\mathrm{T}_{\text{on}} = {h'}/{\mathrm{KL}(f_1,f_0)}$ (when $y_t \sim f_1$) and an off period of $\mathrm{T}_{\text{off}} = {h'}/{\mathrm{KL}(f_0,f_1)}$ (when $y_t \sim f_0$), we have  $0 \leq \rho_t \leq h'$ for $t \geq \tau$.

Since $\rho_t$ is a lower bound for $\mathbb{E}[g_t]$ for $t > 0$, in order to satisfy $\mathbb{E}[g_t] \leq h'$ for $t \geq \tau$, the on period needs to be chosen smaller than ${h'}/{\mathrm{KL}(f_1,f_0)}$ and/or the off period needs to be chosen larger than ${h'}/{\mathrm{KL}(f_0,f_1)}$.


\end{proof}

For a stealthy attack, the attacker needs to choose $h'$ such that $h' < h$. Further, $\Delta \triangleq h - h'$ can be considered as a margin for non-detectability. That is, as $\Delta$ increases, $g_t, t \geq \tau$ takes smaller values on average, that increases the average detection delay of the CUSUM algorithm. Based on Proposition 2, the average on (attacking) period is upper bounded with
\begin{equation}\nonumber
\bar{\mathrm{T}}_{\text{on}} \triangleq \frac{\mathrm{T}_{\text{on}}}{\mathrm{T}_{\text{on}} + \mathrm{T}_{\text{off}}} <
\frac{\mathrm{KL}(f_0,f_1)}{\mathrm{KL}(f_1,f_0) + \mathrm{KL}(f_0,f_1)}.
\end{equation}
Note that the upper bound on $\bar{\mathrm{T}}_{\text{on}}$ is independent of $h'$ and hence of $\Delta$. However, for a higher $\bar{\mathrm{T}}_{\text{on}}$, either $\mathrm{T}_{\text{on}}$ needs to be increased or $\mathrm{T}_{\text{off}}$ needs to be decreased, that both increases $\mathbb{E}[g_t], t \geq \tau$ and hence decreases the average detection delay. Further, note that a stealthy attack can especially be effective if the system has strict false alarm constraints, i.e., requiring high level of false alarm periods and equivalently a high threshold $h$.

\subsubsection{Persistent attacks}

The CUSUM algorithm may not be effective in attack detection if the attack does not comply with the presumed attack model. If an attacker knows that the CUSUM detector is employed based on the post-attack pdf $f_1$, then it can perform a stealthy persistent attack with a post-attack density $f_1' \neq f_1$. The design goal can be keeping $f_1'$ as closest as possible to the post-attack pdf $f_1$ for a strong attack while limiting the risk of being detected. Since the attack is of persistent nature, $f_1'$ needs to be designed such that the decision statistic of the CUSUM algorithm does not increase on average over time. Since the CUSUM algorithm accumulates the LLRs over time (cf. \eqref{eq:cusum}), the LLR can be designed such that it takes non-positive values on the average, i.e., $\mathbb{E}[\ell_t] \leq 0, t > 0$. Then, since $\mathbb{E}[\ell_t] = -\mathrm{KL}(f_0,f_1) < 0$ for $t \leq \tau - 1$, the condition $\mathbb{E}[\ell_t] \leq 0, t \geq \tau$ needs to be satisfied. Considering the KL divergence $\mathrm{KL}(f_1',f_1)$ as the information distance between $f_1'$ and $f_1$, the following optimization problem can be considered:
\begin{align} \label{eq:opt_prob_stealth}
\min_{f_1'} ~ \mathrm{KL}(f_1',f_1) ~~ \text{subject to} ~~ \mathbb{E}[\ell_t] \leq 0, t \geq \tau,
\end{align}
where the solution is presented in the following proposition.

\textbf{Proposition 3:} The solution of \eqref{eq:opt_prob_stealth} is given by
\begin{gather}\label{eq:cond_pers}
\{f_1': \mathrm{KL}(f_1',f_0) = \mathrm{KL}(f_1',f_1)\}.
\end{gather}

\begin{proof}
Let $y_t \sim f_1'$ for $t \geq \tau$. Then,
\begin{align} \nonumber
\mathbb{E}[\ell_t] &= \int {f_1'(y) \log\Big(\frac{f_1(y)}{f_0(y)}\Big) dy} \\  \nonumber
 &= \int {f_1'(y) \log(f_1(y)) dy} - \int {f_1'(y) \log(f_0(y)) dy} \\  \nonumber
 &= \int {f_1'(y) \log(f_1(y)) dy} - \int {f_1'(y) \log(f_1'(y)) dy} + \int {f_1'(y) \log(f_1'(y)) dy} - \int {f_1'(y) \log(f_0(y)) dy} \\  \nonumber
 &= \int {f_1'(y) \log\Big(\frac{f_1(y)}{f_1'(y)}\Big) dy} + \int {f_1'(y) \log\Big(\frac{f_1'(y)}{f_0(y)}\Big) dy} \\  \nonumber
 &= - \mathrm{KL}(f_1',f_1) + \mathrm{KL}(f_1',f_0).
\end{align}
Then, the constraint in \eqref{eq:opt_prob_stealth}, i.e., $\mathbb{E}[\ell_t] \leq 0$,  implies that $- \mathrm{KL}(f_1',f_1) + \mathrm{KL}(f_1',f_0) \leq 0$, which is equivalent to
$\mathrm{KL}(f_1',f_1) \geq \mathrm{KL}(f_1',f_0)$. Hence, the minimum value of $\mathrm{KL}(f_1',f_1)$ is $\mathrm{KL}(f_1',f_0)$.
\end{proof}

Proposition 3 presents a simple strategy for an attacker to perform a persistent stealthy attack against a CUSUM detector. As an example, let $f_0 \sim \mathcal{N}([\mu_0 \, \mu_0]^\mathrm{T},\sigma^2 \, \mathbf{I}_2)$ and $f_1 \sim \mathcal{N}([\mu_1 \, \mu_1]^\mathrm{T},\sigma^2 \, \mathbf{I}_2)$. If
\begin{gather}\nonumber
f_1' \sim \mathcal{N}\bigg(
\begin{bsmallmatrix}
  \frac{1}{2} (\mu_0 + \mu_1) \\
  \frac{1}{2} (\mu_0 + \mu_1)
\end{bsmallmatrix},
\begin{bsmallmatrix}
  \sigma^2 & \varphi \\
  \varphi & \sigma^2
\end{bsmallmatrix}\bigg),
\end{gather}
then it can be checked that
\begin{align} \nonumber
\mathrm{KL}(f_1',f_0) &= \mathrm{KL}(f_1',f_1) \\ \nonumber
&= \frac{1}{4 \sigma^2} (\mu_1 - \mu_0)^2 + \frac{1}{2} \log\Big(\frac{\sigma^4}{\sigma^4 - \varphi^2}\Big),
\end{align}
where the correlation term $\varphi$ can be chosen such that $\sigma^4 - \varphi^2 > 0$.

\subsection{Stealthy Attacks Against Algorithm 1} \label{sec:st_proposed}

In the actual problem under consideration, the pre- and post-attack measurement pdfs, i.e., $p_0(\mathbf{y}_t|\mathbf{x}_t)$ and $p_1(\mathbf{y}_t|\mathbf{x}_t, \mathbf{a}_t, \pmb{\sigma}_t)$, are based on some unknown and time-varying variables and hence the results in the previous subsection do not directly apply. The proposed algorithm estimates the pre- and post-attack pdfs at time $t$ as $p_0(\mathbf{y}_t|\hat{\mathbf{x}}_t^0)$ and $p_1(\mathbf{y}_t|\hat{\mathbf{x}}_t^1, \hat{\mathbf{a}}_t, \hat{\pmb{\sigma}}_t)$, respectively where $\hat{\mathbf{x}}_t^0$ and $\hat{\mathbf{x}}_t^1$ are computed via the Kalman filters, $\hat{\mathbf{a}}_t$ is given in \eqref{eq:a_hat_kt_v3_rrr}, and $\hat{\pmb{\sigma}}_t$ is given in \eqref{eq:sigma_hat_kt_v3_rrr}. Then, the GLLR at time $t$ is computed as follows (cf. \eqref{eq:gen_CUSUM}):
\begin{gather}\label{eq:gllr}
\beta_t = \log\Big(\frac{p_1(\mathbf{y}_t|\hat{\mathbf{x}}_t^1, \hat{\mathbf{a}}_t, \hat{\pmb{\sigma}}_t)}{p_0(\mathbf{y}_t|\hat{\mathbf{x}}_t^0)}\Big).
\end{gather}

Note that computing $\hat{\mathbf{x}}_t^0$ and $\hat{\mathbf{x}}_t^1$ requires the knowledge of all previously taken measurements, i.e., $\{\mathbf{y}_j, j \leq t\}$. Hence, for an attacker, estimating the pre- and post-attack pdfs and hence computing $\beta_t$ and $g_t$ requires monitoring all the system-wide measurements at all times, which is practically infeasible. Although determining the online attack parameters for a stealthy attack is difficult in general, we provide below a brief analysis of the proposed algorithm and discuss possible stealthy attacks against it based on this analysis and the intuitions gained in Sec.~\ref{sec:st_cusum}. As before, since the measurements are random, an attacker can control the decision statistic only on the average.

Firstly, based on \eqref{eq:gllr}, $\beta_t$ depends on how close (relatively) the state estimates $\hat{\mathbf{x}}_t^0$ and $\hat{\mathbf{x}}_t^1$ to the actual system state $\mathbf{x}_t$ and how accurate the MLEs of the attack magnitudes $\hat{\mathbf{a}}_t$ and $\hat{\pmb{\sigma}}_t$ compared to the actual attack magnitudes ${\mathbf{a}}_t$ and ${\pmb{\sigma}}_t$ are. During the pre-attack period, the measurements $\mathbf{y}_t$ follow the normal measurement model \eqref{eq:meas_model} and hence $\mathbf{a}_t = \mathbf{0}$ and $\pmb{\sigma}_t = \mathbf{0}$. Since $\hat{\mathbf{x}}_t^0$ is computed assuming no attack, for $t \leq \tau-1$, $\hat{\mathbf{x}}_t^0$ is usually a better estimate of the actual system state $\mathbf{x}_t$ compared to $\hat{\mathbf{x}}_t^1$ due to the possible ML estimation errors in computing $\hat{\mathbf{a}}_t$, $\hat{\pmb{\sigma}}_t$, and $\hat{\mathbf{x}}_t^1$ (recall that (cf. \eqref{eq:meas_upd_fdata_alter}) $\hat{\mathbf{x}}_t^1$ is computed based on $\hat{\mathbf{a}}_t$ and $\hat{\pmb{\sigma}}_t$). Then, $p_0(\mathbf{y}_t|\hat{\mathbf{x}}_t^0)$ is expected to fit better to $\mathbf{y}_t$ compared to $p_1(\mathbf{y}_t|\hat{\mathbf{x}}_t^1, \hat{\mathbf{a}}_t, \hat{\pmb{\sigma}}_t)$, i.e., we usually have $p_0(\mathbf{y}_t|\hat{\mathbf{x}}_t^0) \geq p_1(\mathbf{y}_t|\hat{\mathbf{x}}_t^1, \hat{\mathbf{a}}_t, \hat{\pmb{\sigma}}_t)$ for $t \leq \tau-1$. Then, based on \eqref{eq:gllr}, $\beta_t$ is expected to take nonpositive values in general, that makes $g_t \approx 0$ a good approximation for $t \leq \tau-1$.

The main aim of an attacker against the smart grid is deviating the state estimates from the actual system state as much as possible without being detected. Hence, it needs to keep $g_t$ below the level of $h$ as long as possible for $t \geq \tau$ (cf. \eqref{eq:gen_CUSUM}). Similar to the stealthy attacks against the CUSUM algorithm discussed in Sec.~\ref{sec:st_cusum}, an attacker can either follow an on-off attacking strategy or perform persistent attacks that do not comply with the presumed attack magnitudes (cf. \eqref{eq:change_event}). Note that in case of an attack, i.e., for $t \geq \tau$, $\hat{\mathbf{x}}_t^0$ deviates from $\mathbf{x}_t$ since it is computed assuming no attack. On the other hand, $\hat{\mathbf{x}}_t^1$ is a recovered state estimate, but it is subject to possible ML estimation errors.


\subsubsection{Non-persistent attacks}

Since Algorithm 1 is a CUSUM-based detector, stealthy intermittent (on-off) attacking can be performed against it. Specifically, during the on periods, an attacker can choose its attack magnitudes comparable to or larger than the presumed lower bounds on the attack magnitudes, i.e., $\gamma$ and $\sigma^2$, with the purpose of a strong attack. However, as explained before, analytically deriving the on-off periods and the online attack magnitudes seems infeasible for an attacker. Hence, a smart attacker, having the knowledge of system and detector parameters, can determine its attack parameters based on an offline simulation.

When the attack complies with the presumed attack magnitudes, during an on period, $\hat{\mathbf{x}}_t^0$ usually deviates from the actual system state more than $\hat{\mathbf{x}}_t^1$ due to the recovery mechanism in computing $\hat{\mathbf{x}}_t^1$ (cf. \eqref{eq:meas_upd_fdata_alter}). That makes $p_1(\mathbf{y}_t|\hat{\mathbf{x}}_t^1, \hat{\mathbf{a}}_t, \hat{\pmb{\sigma}}_t)$ a better fit to $\mathbf{y}_t$ compared to $p_0(\mathbf{y}_t|\hat{\mathbf{x}}_t^0)$ on the average. Then, if an on-off attack is performed, during the on periods, based on \eqref{eq:gllr}, usually $\beta_t$ takes non-negative values and $g_t$ increases. Further, at the beginning of an off period after an on period, although the attack magnitudes are zero, i.e., $\mathbf{a}_t = \pmb{\sigma}_t = \mathbf{0}$, since $\hat{\mathbf{x}}_t^1$ is still a better (recovered) state estimate compared to $\hat{\mathbf{x}}_t^0$, $p_1(\mathbf{y}_t|\hat{\mathbf{x}}_t^1, \hat{\mathbf{a}}_t, \hat{\pmb{\sigma}}_t)$ can still be a better fit to $\mathbf{y}_t$ and $g_t$ may further increase. Note that during an off period, $\hat{\mathbf{x}}_t^0$ is not expected to deviate further since $\mathbf{y}_t$ now follows the normal measurement model. On the other hand, $\hat{\mathbf{a}}_t$, $\hat{\pmb{\sigma}}_t$, and $\hat{\mathbf{x}}_t^1$ are still subject to possible ML estimation errors. That may make $p_0(\mathbf{y}_t|\hat{\mathbf{x}}_t^0)$ a better fit to $\mathbf{y}_t$ over time and as the off period is continued, $\beta_t$ may start to take nonpositive values on the average. Since $g_t$ is expected to increase in an on period as well as in the beginning of an off period, the level of attack magnitudes during the on periods needs to be carefully chosen in accordance with the aim of keeping the highest value of $g_t$ below the decision threshold $h$, for $t \geq \tau$.

\subsubsection{Persistent attacks}

Since Algorithm 1 relies on the lower bounds $\gamma$ and $\sigma^2$ defined on the attack magnitudes (cf. \eqref{eq:change_event}), an attacker can perform persistent stealthy attacks using significantly small attack magnitudes compared to $\gamma$ and $\sigma^2$ so that Algorithm 1 becomes ineffective to detect such attacks. In case of such small-magnitude stealthy attacks, the attack magnitudes ${\mathbf{a}}_t$ and ${\pmb{\sigma}}_t$ are close to zero for $t \geq \tau$ and due to the possible ML estimation errors in computing $\hat{\mathbf{a}}_t$ and $\hat{\pmb{\sigma}}_t$, $\hat{\mathbf{x}}_t^1$ usually deviates more compared to $\hat{\mathbf{x}}_t^0$, similarly to the pre-attack period discussed before. Then, $p_0(\mathbf{y}_t|\hat{\mathbf{x}}_t^0)$ fits better to $\mathbf{y}_t$ compared to $p_1(\mathbf{y}_t|\hat{\mathbf{x}}_t^1, \hat{\mathbf{a}}_t, \hat{\pmb{\sigma}}_t)$ on the average. Hence, in case of a persistent small-magnitude attack, based on \eqref{eq:gllr}, $\beta_t$ usually takes nonpositive values and the approximation $g_t \approx 0$ can still be valid. Note that even if an attacker has an incomplete knowledge about the system and detector parameters, it can still perform stealthy attacks with small attack magnitudes. Although such small-magnitude attacks have minimal effects on the system performance in the short run, they can be effective over long periods of time. Hence, they need to be detected with reasonable detection delays.


\subsection{Countermeasures Against Stealthy Attacks} \label{sec:st_counter}

We firstly discuss countermeasures against the on-off attacking strategy, i.e., the non-persistent stealthy attacks. We then discuss a countermeasure against the persistent stealthy attacks where the attacks may not comply with the presumed attack model/magnitudes. Finally, we propose a new detection scheme, i.e., Algorithm \ref{alg:countermeasure}, that simultaneously employs Algorithm \ref{alg:centralized} and the proposed countermeasures with the aim of being effective against a diverse range of potential cyber-attacks.

\subsubsection{Countermeasures against the non-persistent stealthy attacks} \label{sec:non_persistent}

Timely detection of cyber-attacks against the smart grid is crucial since any failure may quickly spread over the network. Hence, in practice, detection delays cannot be allowed to be arbitrarily large. Note that the optimization problem stated in \eqref{eq:opt_prob} does not impose any constraints on the maximum tolerable detection delay. In an alternative quickest detection problem considered in \cite{Moustakides14}, the objective is maximizing the probability of detection in at most $\eta$ time units after an attack occurs, where $\tau \leq T < \tau+\eta$ needs to be satisfied for a successful detection.

In the extreme case of the considered non-persistent stealthy attacks, the attacker can choose its on period as $\mathrm{T}_{\text{on}} = 1$. In such a case, the attack needs to be detected using the measurements obtained during a single time interval in the attacking regime and hence $\eta = 1$ needs to be chosen. Then, we consider the following optimization problem, proposed in \cite{Moustakides14}:
\begin{gather}\label{eq:opt_shewhart}
\sup_T ~ p(T) ~~ \text{subject to} ~~ \mathbb{E}_\infty[T] \geq \alpha,
\end{gather}
where
\begin{gather}\label{eq:worst_prob} \nonumber
p(T) = \inf_{\tau} \, \essinf_{\mathcal{F}_\tau} \, \mathbb{P}_\tau \big(T = \tau\,|\mathcal{F}_\tau, T\geq\tau\,\big)
\end{gather}
is the worst-case (in the Lorden's sense) detection probability after obtaining the first measurements in the attacking regime.

\vspace{0.2cm}

\textbf{Shewhart Test}: In case the pre- and post-attack pdfs are known as in Sec.~\ref{sec:st_cusum}, the optimum solution to \eqref{eq:opt_shewhart} is the Shewhart test \cite[Theorem 2.3]{Moustakides14}, given by
\begin{gather}\label{eq:Shewhart}
T_S = \inf \{t: \ell_t \geq \nu\},
\end{gather}
where $T_S$ denotes the stopping time and the threshold $\nu$ is determined such that $\mathbb{P}_{\infty}(\ell_1 \geq \nu) = {1}/{\alpha}$. Note that the Shewhart test in \eqref{eq:Shewhart} is, in fact, the repeated log-likelihood ratio test (LLRT), i.e., at each time, the LLR is compared with a certain threshold and an alarm is triggered at the first time the LLR crosses the threshold. We then propose the Shewhart test as a countermeasure to the non-persistent stealthy attacks against the CUSUM detector. Note that $\nu$ needs to be chosen sufficiently high to prevent frequent false alarms.

\vspace{0.2cm}

\textbf{Generalized Shewhart Test}: In case the pre- and post-attack pdfs are unknown and time-varying, we can only compute the GLLR $\beta_t$. Hence, as a countermeasure, we propose to employ the generalized Shewhart test, i.e., the repeated generalized LLRT (GLLRT), given by
\begin{gather}\label{GLLRT}
T' = \inf\{t: \beta_t \geq \phi\},
\end{gather}
where $T'$ is the stopping time and $\phi$ is the test threshold. Again, a sufficiently high threshold needs to be chosen to prevent frequent false alarms. Moreover, similar to Algorithm 1, by choosing higher $\gamma$ and/or $\sigma^2$, false alarm rate of the generalized Shewhart test can be reduced (see Remark 2).

Note that the generalized Shewhart test is expected to be mainly effective in detecting significant instantaneous increases in the level of GLLR and hence in detecting non-persistent stealthy attacks during the on periods. That is, even if an attack may be missed by Algorithm 1 since the decision statistic $g_t$ may not achieve reliably high values for $t \geq \tau$ due to the subsequent off period after an on period, the generalized Shewhart test can detect such non-persistent increases during the on periods.

\subsubsection{A countermeasure against the persistent stealthy attacks} \label{subsec:counter}

Non-parametric detection techniques do not assume any attack models and only evaluate the deviation of measurement statistics from the baseline (no attack) statistics. However, they are usually less effective if the attacks comply with the presumed attack models. As explained before, a persistent stealthy attack can be performed if an attack does not match the considered attack models or magnitudes. For such cases, parametric detectors such as Algorithm 1 and the generalized Shewhart test become ineffective and the non-parametric detection techniques become more appropriate.

In case of no attack, i.e., for $t < \tau$, $c_t \triangleq {\mathbf{r}_t}^T  {\mathbf{Q}_t}^{-1} \mathbf{r}_t$ is a chi-square random variable with $K \lambda$ degrees of freedom \cite{Brumback87}, where $\mathbf{r}_t$ denotes the measurement innovation signal at time $t$, given as
\begin{gather} \label{eq:residual} \nonumber
\mathbf{r}_t \triangleq \mathbf{y}_t - \mathbf{H} \hat{\mathbf{x}}_{t|t-1}^{0},
\end{gather}
and $\mathbf{Q}_t$ is the measurement prediction covariance matrix at time $t$, calculated as follows:
\begin{gather} \label{eq:meas_pred_cov} \nonumber
\mathbf{Q}_t \triangleq \mathbf{H} \mathbf{P}_{t|t-1} \mathbf{H}^T + \sigma_w^2 \, \mathbf{I}_{K \lambda}.
\end{gather}
Notice that although the distribution of the measurements $\mathbf{y}_t$ is time-varying and unknown due to the dynamic system state $\mathbf{x}_t$, the distribution of $c_t$ is time-invariant and known in case of no attack, i.e., for $t < \tau$. Hence, whether the sequence $\{c_t\}$ fits to the chi-squared distribution or not can be evaluated via a goodness-of-fit test and if not, an attack/anomaly is declared.

\vspace{0.2cm}

\textbf{Sliding-Window Chi-Squared Test}: We partition the range of $c_t$, i.e., $[0, \infty)$, into $M$ mutually exclusive and disjoint intervals $I_j, j = 1,2, \dots, M$ such that $p_1 = P(c_t \in I_1)$, $p_2 = P(c_t \in I_2)$, \dots, $p_M = P(c_t \in I_M)$. Hence, $p_1, p_2, \dots, p_M$ denote the probabilities that $c_t$ belongs to the intervals $I_1, I_2, \dots, I_M$, respectively for $t < \tau$, where $\sum_{j=1}^{M} p_j = 1$. The intervals $I_1, I_2, \dots, I_M$ can be determined using the cumulative distribution function (cdf) of a chi-squared random variable with $K \lambda$ degrees of freedom. Then, the null hypothesis is that $c_t$ belongs to the intervals $I_1, I_2, \dots, I_M$ with probabilities $p_1, p_2, \dots, p_M$, respectively.

We propose to employ an online test to evaluate whether the most recent sliding window of $c_t$'s fits to the null hypothesis. Let the size of the sliding window be $L$. Then, the sliding window at time $t$, denoted with $\mathbf{W}_t$, consists of $\{c_j: t-L+1 \leq j \leq t\}$. Let the number of samples in $\mathbf{W}_t$ belonging to the predetermined disjoint intervals be denoted with $N_{1,t}, N_{2,t}, \dots, N_{M,t}$, respectively where $\sum_{i=1}^{M} N_{i,t} = L, \forall t$. Since we have a multinomial distribution where the expected number of samples in the disjoint intervals are $L p_1, L p_2, \dots, L p_M$, respectively, the Pearson's chi-squared test can be used to evaluate the goodness of fit, that can be written as
\begin{gather}\label{chi_squ}
T'' = \inf\{t: \chi_t \geq \varphi\},
\end{gather}
where
\begin{gather}\label{eq:Pearson}
\chi_t = \sum_{i=1}^{M} \frac{(N_{i,t} - L p_i)^2}{L p_i}
\end{gather}
is the asymptotically (as $L \rightarrow \infty$) chi-squared distributed test statistic with $M-1$ degrees of freedom under the null hypothesis, $\varphi$ is the test threshold that can be determined using the cdf of a chi-squared random variable for a desired significance level, and $T''$ denotes the stopping time. To improve the accuracy of the detector, $M$ can be chosen higher. Note, however, that as $M$ is increased, the window size $L$ needs also to be increased to improve the reliability of the goodness-of-fit test, that will cause larger detection delays.

{The chi-squared test does not assume any attack model a priori and it only evaluates deviation of observed measurements from the baseline statistics corresponding to the normal system operation. We propose to use the chi-squared test to have a detection scheme that is robust
against (i) low-magnitude stealthy attacks that corresponds to small deviations from the baseline for which our proposed parametric detectors (Algorithm 1 and the generalized Shewhart test) become ineffective to detect, and (ii) attacks that do not comply with the presumed hybrid attack model, e.g., non-Gaussian or correlated jamming noise.}

\textit{Remark 3:} The proposed chi-squared test is a sequential version of the Pearson's chi-squared test since the most recent sliding window of samples is used in the test. Furthermore, the proposed test is different from the outlier detector in \cite{Brumback87}, that makes sample by sample decisions, i.e., it declares a single sample as either normal and anomalous. The proposed test is thus more reliable and more sensitive to small deviations from the baseline since it considers long-term deviations by evaluating a sliding window of samples.

\subsubsection{Proposed final detection scheme}

Our aim is to obtain a detection mechanism that is effective against a significantly wide range of cyber-attacks. Hence, we propose to simultaneously employ Algorithm 1, the generalized Shewhart test, and the sliding-window chi-squared test and declare an attack at the first time instant one of the detectors declares an attack (if any). Hence,
\begin{gather}\nonumber
\tilde{T} = \inf\{T,T',T''\}
\end{gather}
is the proposed stopping time. We summarize the proposed detector in Algorithm \ref{alg:countermeasure}. Note that the average false alarm period of Algorithm \ref{alg:countermeasure} is less than the minimum of the (individual) average false alarm periods of Algorithm 1, the generalized Shewhart test, and the sliding-window chi-squared test. Hence, sufficiently high thresholds, i.e., $h$, $\phi$ and $\varphi$, need to be chosen to prevent frequent false alarms. Furthermore, to have the same average false alarm periods $\alpha$ for Algorithms 1 and 2, the threshold $h$ needs to be chosen higher in Algorithm 2 and the thresholds $\phi$ and $\varphi$ need to be chosen such that the individual average false alarm periods of the generalized Shewhart test and the sliding-window chi-squared tests are greater than $\alpha$.

\begin{algorithm}[t]\small
\caption{\small Real-time detection of hybrid and stealthy attacks}
\label{alg:countermeasure}
\baselineskip=0.4cm
\begin{algorithmic}[1]
\STATE Initialization: $t \gets 0$, $g_{0} \gets 0$, $\hat{\tau} \gets 1$, choose the entries of the initial sliding window of the chi-squared test, i.e., $\mathbf{W}_0$, as realizations of a chi-squared random variable with $K \lambda$ degrees of freedom.
\WHILE {$g_t < h$ and $\beta_t < \phi$ and $\chi_t < \varphi$}
    \STATE $t \gets t+1$
    \STATE Implement the lines 4--15 in Algorithm 1.
    \STATE $\mathbf{r}_t \gets \mathbf{y}_t - \mathbf{H} \hat{\mathbf{x}}_{t|t-1}^{0}$
    \STATE $\mathbf{Q}_t \gets \mathbf{H} \mathbf{P}_{t|t-1} \mathbf{H}^T + \sigma_w^2 \, \mathbf{I}_{K \lambda}$
    \STATE $c_t \gets {\mathbf{r}_t}^T  {\mathbf{Q}_t}^{-1} \mathbf{r}_t$
    \STATE Update $\mathbf{W}_t$ with the most recent entry $c_t$.
    \STATE Update $N_{1,t}, N_{2,t}, \dots, N_{M,t}$ and compute $\chi_t$ using \eqref{eq:Pearson}.
\ENDWHILE
\STATE $\tilde{T} \gets t$, declare a cyber-attack.
\end{algorithmic}
\end{algorithm}

\section{Simulation Results} \label{sec:numerical}

In this section, we evaluate the performance of the proposed detection schemes via simple case studies in an IEEE-14 bus power system, where $K = 23$ and $N = 13$. The system matrix $\mathbf{A}$ is chosen to be an identity matrix, the measurement matrix $\mathbf{H}$ is determined based on the IEEE-14 bus power system, and $\lambda = 5$ is chosen. The initial state variables are obtained through the DC optimal power flow algorithm for case-14 in MATPOWER \cite{Zimmerman11}. The system noise variances are chosen as $\sigma_v^2 = \sigma_w^2 = 10^{-4}$. Furthermore, the parameters of Algorithm 1 are chosen as $\gamma = 0.022$ and $\sigma^2 = 10^{-2}$. In Algorithm 2, the threshold of the generalized Shewhart test is chosen as $\phi = 10$. Moreover, for the chi-squared test, the window size is chosen as $L = 80$, number of disjoint intervals are chosen as $M = 5$, the probabilities are chosen as $p_1=p_2=\dots=p_5 = 0.2$, and the intervals $I_1, I_2, \dots, I_5$ are determined accordingly as $I_1 = [0,102.081)$, $I_2 = [102.081,110.5475)$, $I_3 = [110.5475,118.2061)$, $I_4 = [118.2061,127.531)$, and $I_5 = [127.531,\infty)$ based on the cdf of a chi-squared random variable with $K \lambda = 115$ degrees of freedom. The threshold of the Pearson's chi-squared test, i.e., $\varphi = 25.0133$, is chosen based on the significance level of $5 \times 10^{-5}$ for a chi-squared random variable with $M-1 = 4$ degrees of freedom. The thresholds of the generalized Shewhart and the chi-squared tests are chosen such that the (individual) average false alarm periods of these tests are in the order of $10^4$. The cyber-attacks are launched at $t = 100$.

Firstly, we evaluate the performance of the proposed schemes in case of an FDI-only attack, a jamming-only attack, and a hybrid attack. We then evaluate the performance in case of stealthy hybrid attacks. Particularly, we consider a non-persistent stealthy attack and a small-magnitude persistent stealthy attack, and illustrate the performance improvement obtained with the proposed countermeasures against such stealthy attacks. Next, we illustrate the mean squared error (MSE) vs. time plot for the recovered and non-recovered state estimates in case of a hybrid cyber-attack.  {Finally, we evaluate the performance of the proposed schemes in case of a network topology attack/failure.}

\subsection{Case 1: FDI Attack} \label{sec:num_fdi}

\begin{figure}
\center
  \includegraphics[width=78mm]{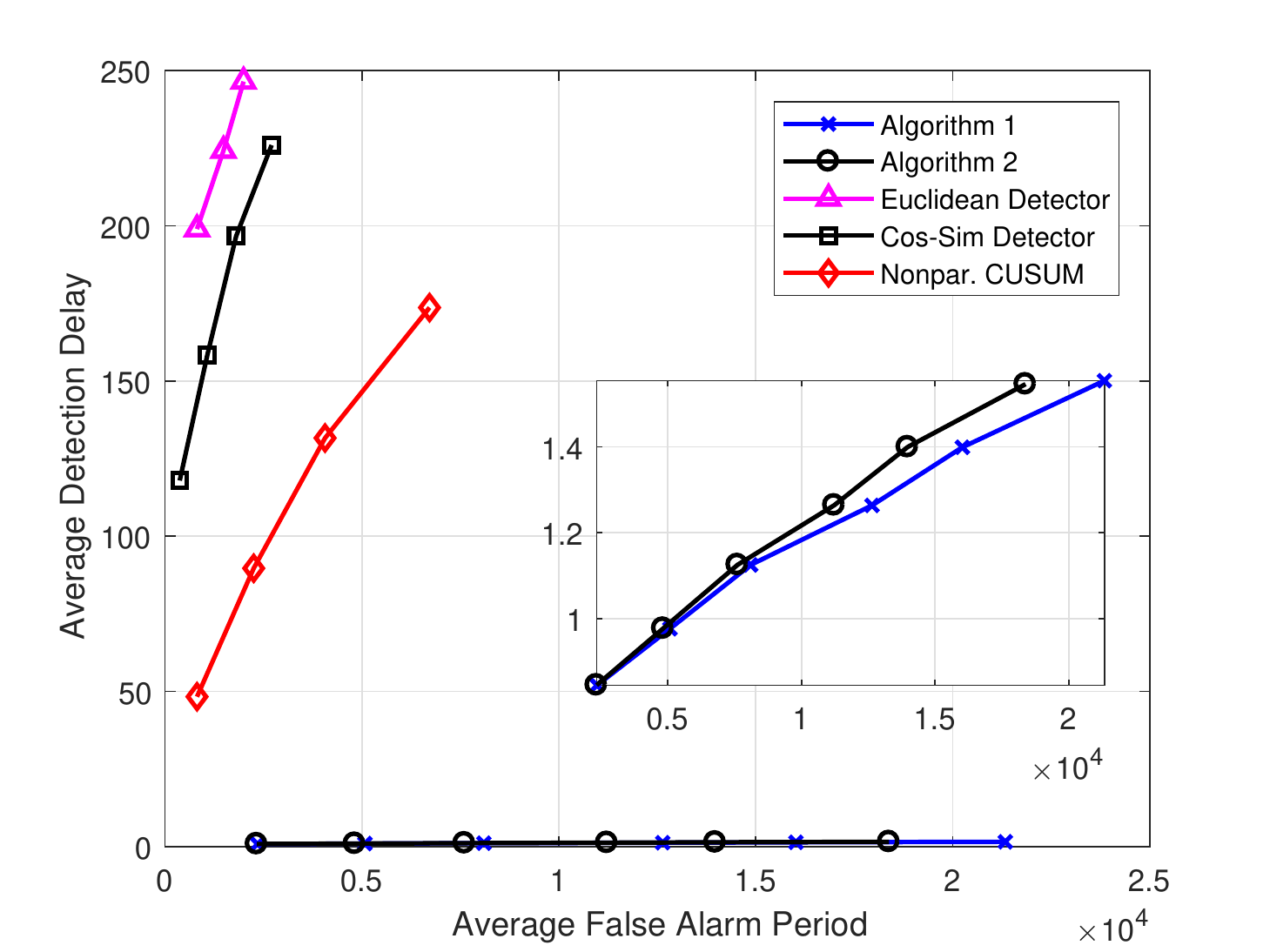}
\caption{Average detection delay vs. average false alarm period for the proposed detectors and the benchmark tests in case of a random FDI attack.}
 \label{fig:fdata_tradeoff}
\end{figure}

We firstly consider a random and time-varying persistent FDI attack where at each time the attacker chooses the magnitudes of the injected false data and the set of attacked meters randomly. In particular, at each time, the attacker compromises the measurements of each meter with probability $0.5$ and injects the realizations of the uniform random variable $\mathcal{U}[-0.02,0.02]$ to the attacked meters. Fig.~\ref{fig:fdata_tradeoff} illustrates the tradeoff between the average detection delay and the average false alarm period for the proposed algorithms and also {three benchmark tests, namely the nonparametric CUSUM test in \cite{yang2016false},} the Euclidean detector \cite{Manandhar14} and the cosine-similarity metric based detector \cite{Rawat15} that both check the dissimilarity between the actual and the predicted measurements (by the Kalman filter) and declare an anomaly if the dissimilarity metric is greater than certain thresholds.

{Since a nonlinear power system model is studied in \cite{yang2016false} and we use a linear system model, we include a modified version of the nonparametric CUSUM detector for the linear case. The stopping time and the update of the decision statistic over time for the modified nonparametric CUSUM detector are given as follows:
\begin{align} \nonumber
\bar{T} &\triangleq \inf\left\{t: S_t \geq q \right\}, \\ \nonumber
S_t &= S_{t-1} + \delta_t, \\ \nonumber
\delta_t &\triangleq \|\mathbf{y}_t - \mathbf{H} \hat{\mathbf{x}}_{t|t-1}^0\| - \mathbb{E}_0\big[\|\mathbf{y}_t - \mathbf{H} \hat{\mathbf{x}}_{t|t-1}^0\|\big],
\end{align}
where $\bar{T}$ denotes the corresponding stopping time, $S_t$ is the decision statistic at time $t$ where $S_0 = 0$, $q$ is the test threshold that controls the false alarm rate of the detector, and $\mathbb{E}_0[\|\mathbf{y}_t - \mathbf{H} \hat{\mathbf{x}}_{t|t-1}^0\|]$ denotes the expectation of the $L_2$ norm of the measurement innovation signal in the pre-change case, computed via a Monte Carlo simulation. The nonparametric CUSUM detector accumulates the difference between magnitude of the measurement innovation signal and its expected value in the pre-change case (normal system operation).}

We observe that the proposed algorithms significantly outperform the benchmark tests. Moreover, Algorithm 1 slightly outperforms Algorithm 2. This is because the countermeasures introduced in Algorithm 2 slightly increase the false alarm rate of Algorithm 2. Note that in obtaining the tradeoff curve for Algorithm 2, we keep the thresholds $\phi$ and $\varphi$ constant and only vary $h$.


\begin{figure}
\center
  \includegraphics[width=78mm]{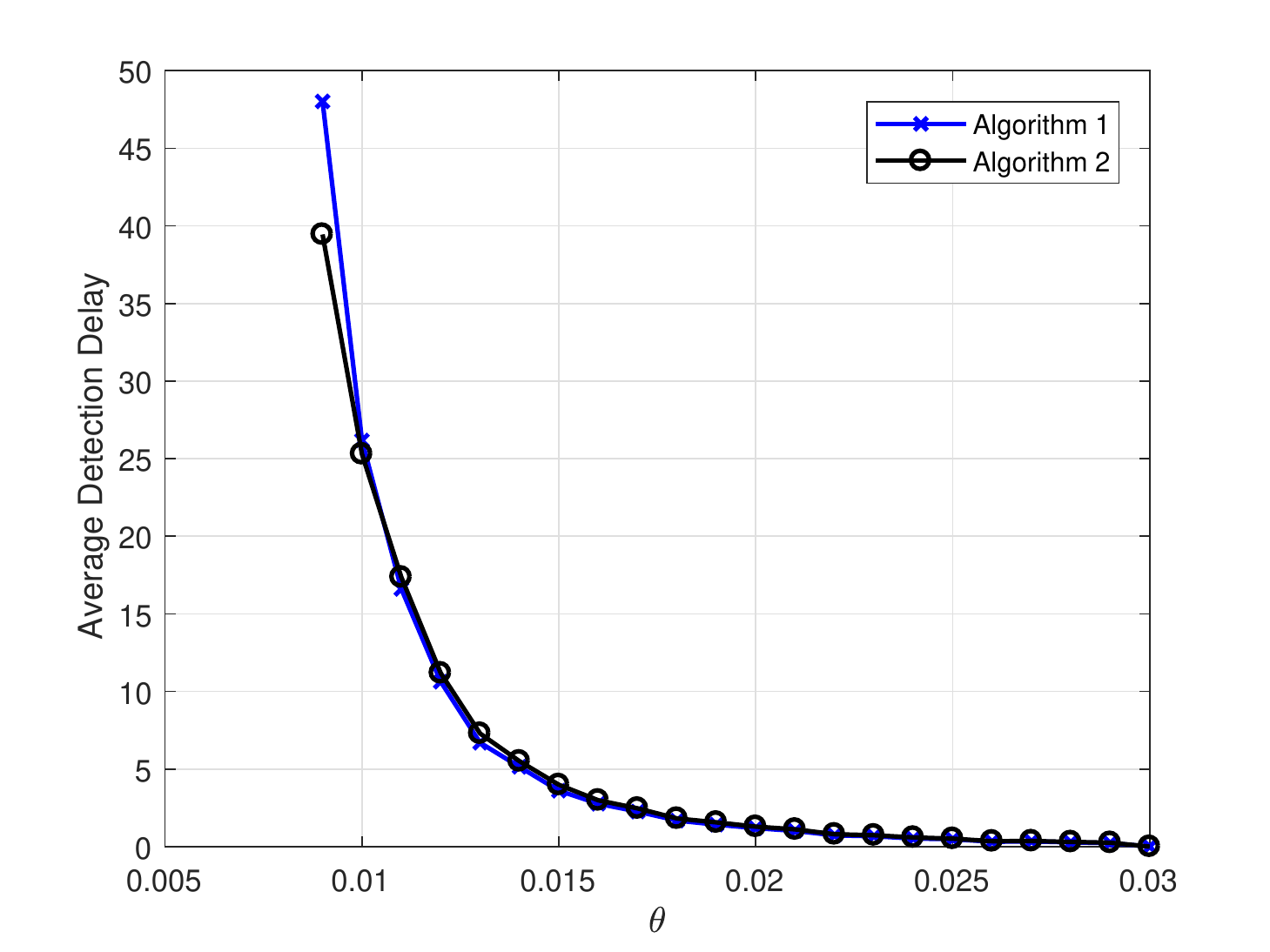}
  \vspace{-0.1cm}
\caption{Average detection delay vs. magnitude of the injected false data for the proposed detectors in case of a random FDI attack, where the average false alarm period is approximately $1.5\times10^{4}$.}
 \label{fig:fdata_add_vs_mag}
\end{figure}

We then illustrate the performance of the proposed algorithms as the magnitude of the injected false data varies while keeping the false alarm rate constant. We again consider the random and time-varying persistent FDI attack described above, but this time the magnitudes of the injected data are realizations of $\mathcal{U}[-\theta,\theta]$, where $\theta$ varies between $0.009$ and $0.03$. Through Fig.~\ref{fig:fdata_add_vs_mag}, we see the advantage of the proposed countermeasures as the magnitude of the false data takes very small values. For instance, when $\theta = 0.009$, the average detection delays of Algorithm 1 and Algorithm 2 are $48.02$ and $39.45$, respectively.

\subsection{Case 2: Jamming Attack} \label{sec:num_jam}

\begin{figure}
\center
  \includegraphics[width=78mm]{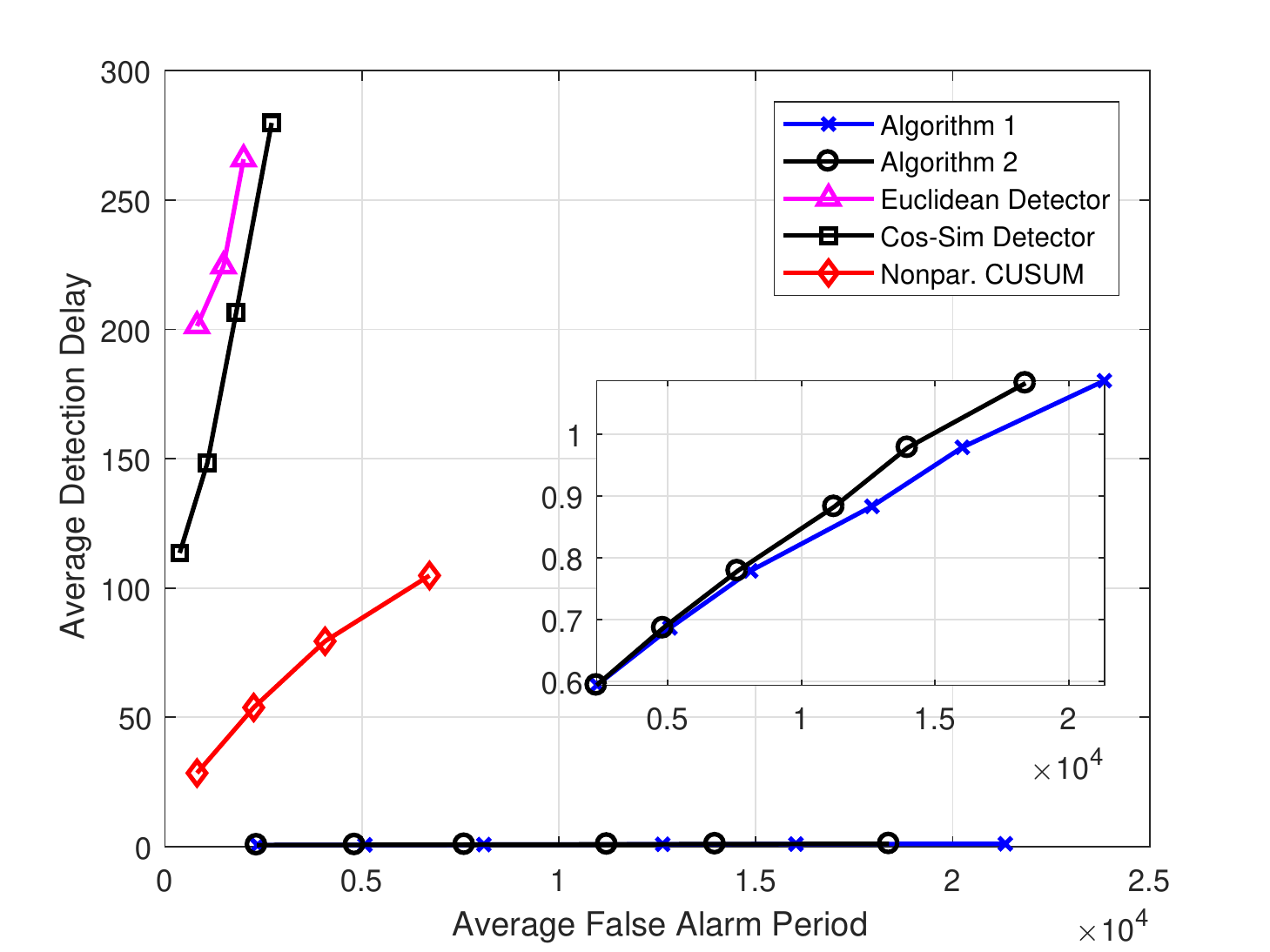}
\caption{Average detection delay vs. average false alarm period for the proposed detectors and the benchmark tests in case of a random jamming attack.}
 \label{fig:jamming_tradeoff}
\end{figure}

\begin{figure}
\center
  \includegraphics[width=78mm]{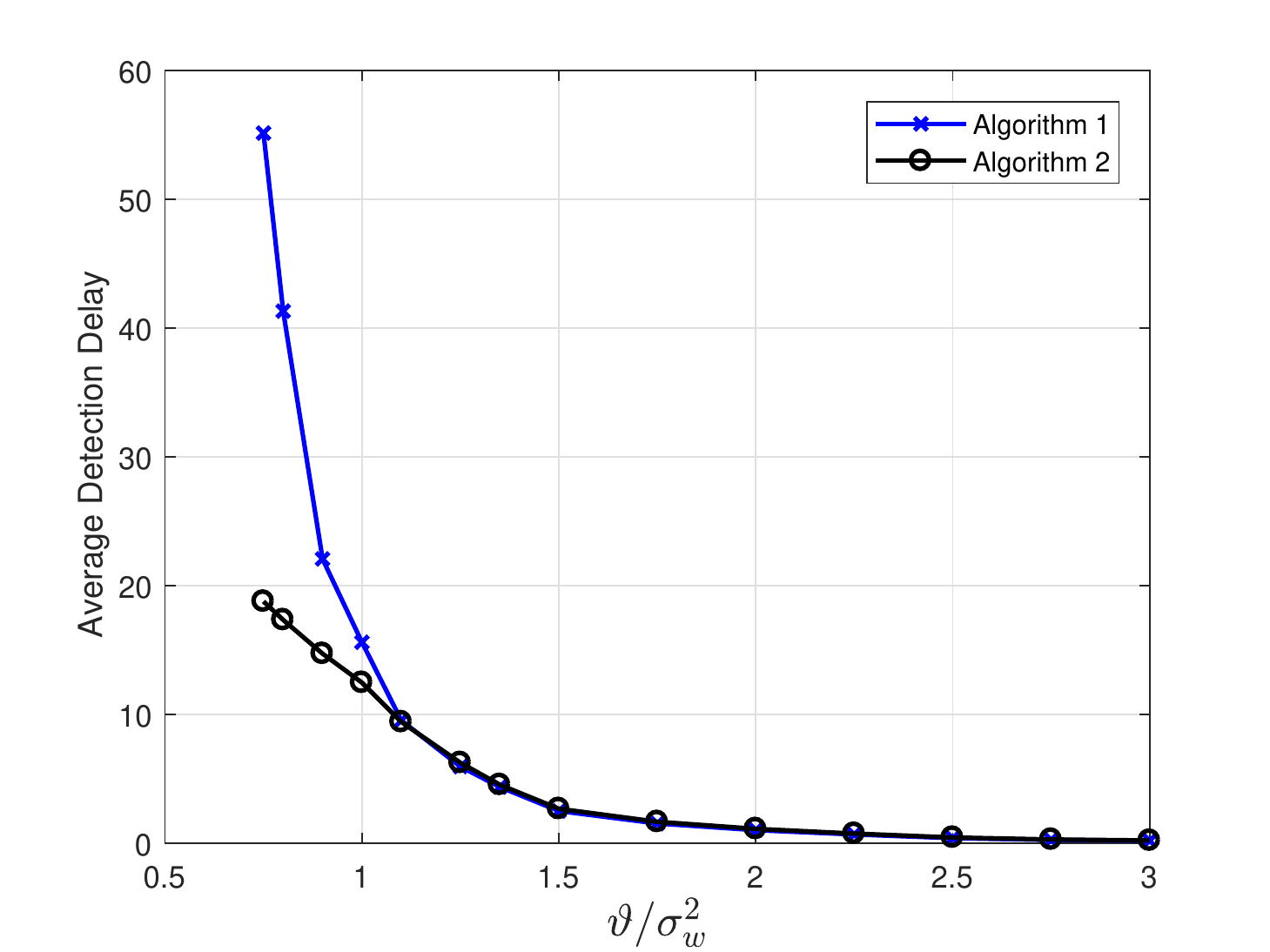}
  \vspace{-0.1cm}
\caption{Average detection delay vs. variance of the jamming noise for the proposed detectors in case of a random jamming attack, where the average false alarm period is approximately $1.5\times10^{4}$.}
 \label{fig:jamming_add_vs_mag}
\end{figure}

Next, we consider a random and time-varying persistent jamming attack. At each time, an attacker jams the measurements of each meter with probability $0.5$ where the variances of the jamming noise are the realizations of the uniform random variable $\mathcal{U}[2\times10^{-4},4\times10^{-4}]$. Fig.~\ref{fig:jamming_tradeoff} presents the delay to false alarm curve for the proposed algorithms and the benchmark tests. Further, we evaluate the performance as the magnitude of the jamming noise variance varies by keeping the false alarm rate constant. In particular, jamming noise variances are chosen as realizations of $\mathcal{U}[\vartheta,2 \vartheta]$, where $\vartheta$ is varied between $0.75 \sigma_w^2$ and $3 \sigma_w^2$. Through Fig.~\ref{fig:jamming_add_vs_mag}, we again observe smaller detection delays in Algorithm 2 compared to Algorithm 1 in case of very small attack magnitudes.

\subsection{Case 3: Hybrid FDI/Jamming Attack} \label{sec:num_combined}

\begin{figure}
\center
  \includegraphics[width=78mm]{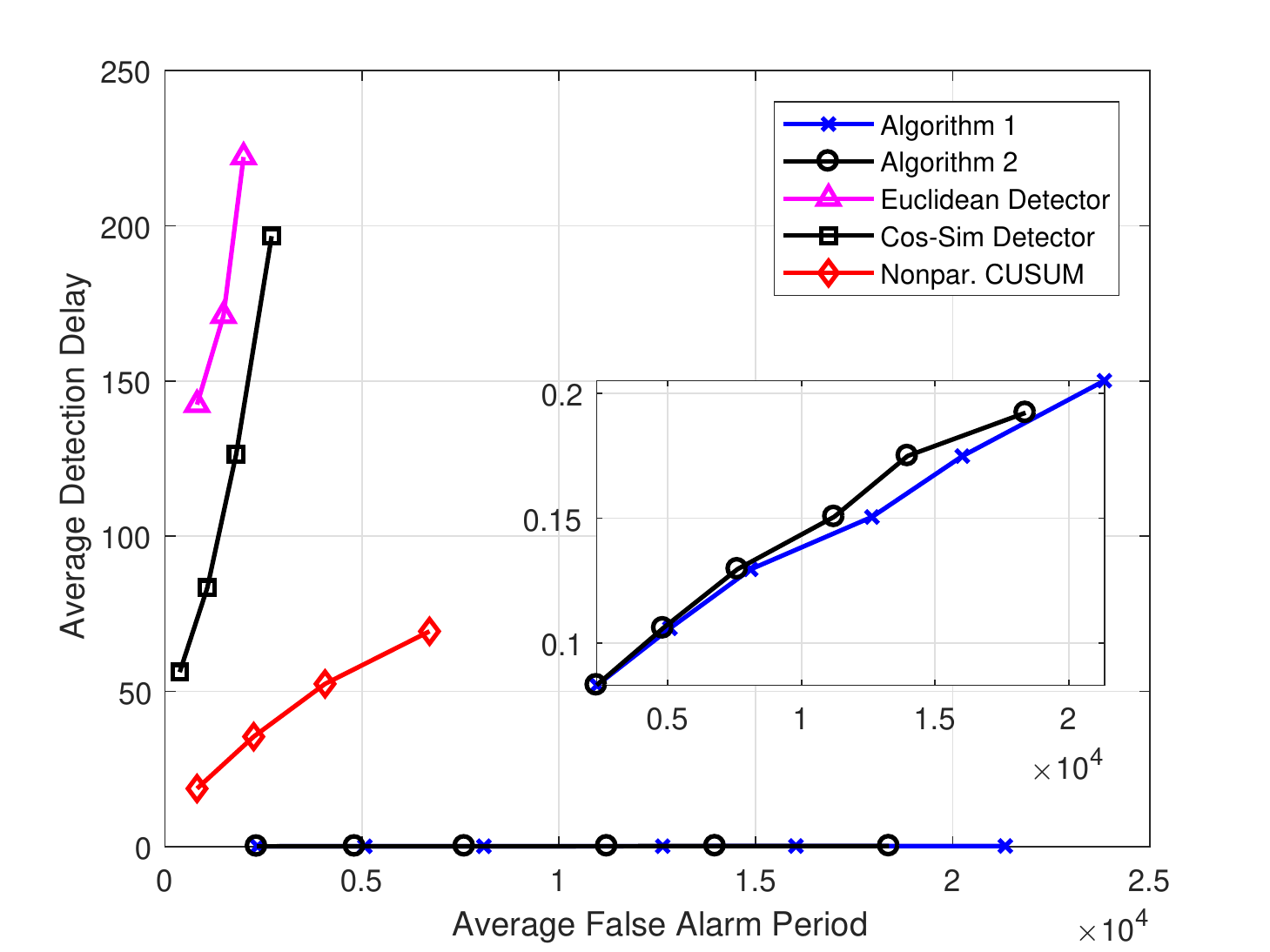}
\caption{Average detection delay vs. average false alarm period for the proposed detectors and the benchmark tests in case of a hybrid attack.}
 \label{fig:combined_tradeoff}
\end{figure}

Next, we consider a random and time-varying persistent hybrid attack. The attack is combined over the system and it may also be combined over a subset of meters. In particular, we consider the attacks described in Sec.~\ref{sec:num_fdi} and Sec.~\ref{sec:num_jam} altogether. Hence, the attacker chooses a random subset of meters for FDI attack and another random subset of meters for jamming attack, where these subsets might overlap with each other. The attack magnitudes for FDI and jamming attacks are realizations of $\mathcal{U}[-0.02,0.02]$ and $\mathcal{U}[2\times10^{-4},4\times10^{-4}]$, respectively. In Fig.~\ref{fig:combined_tradeoff}, for the same levels of false alarm rate, we observe smaller detection delays compared to Figures \ref{fig:fdata_tradeoff} and \ref{fig:jamming_tradeoff}, as expected.

\subsection{Case 4: Non-persistent Stealthy Attack} \label{sec:nonper_st}

\begin{figure}
\center
  \includegraphics[width=78mm]{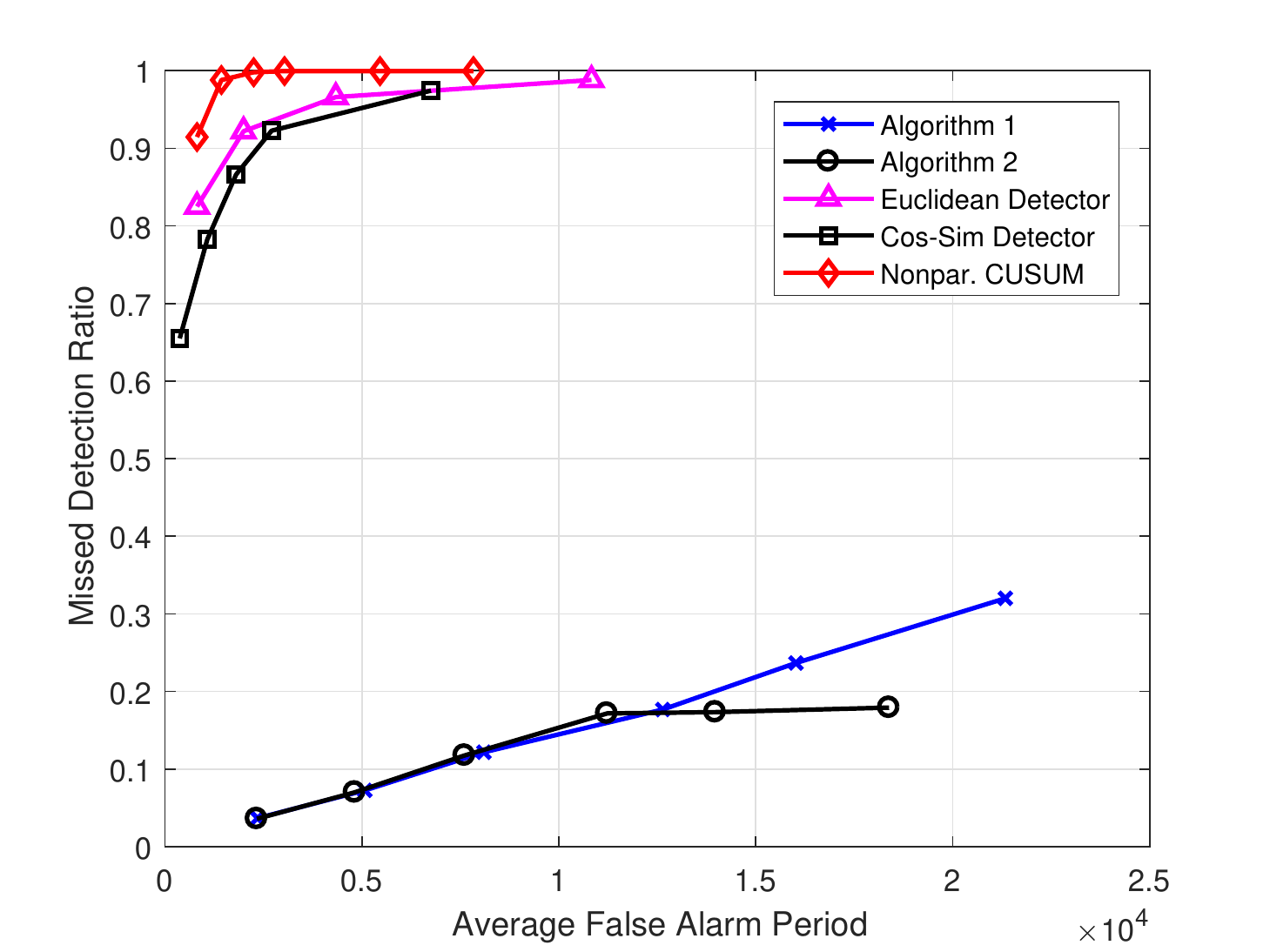}
\caption{Missed detection ratio vs. average false alarm period for the proposed detectors and the benchmark tests in case of a stealthy non-persistent attack, where the attack is assumed to be missed if it is not detected within $50$ time units.}
 \label{fig:miss_non_persistent}
\end{figure}

Next, we consider a stealthily designed on-off attack. Particularly, after the attack is launched at $t = 100$, the attacker performs a hybrid attack as described in Sec.~\ref{sec:num_combined} where the magnitudes of the FDI and jamming attacks are realizations of $\mathcal{U}[-0.01,0.01]$ and $\mathcal{U}[10^{-4},2\times10^{-4}]$, respectively and the on and off periods are $\mathrm{T}_{\text{on}} = 1$ and $\mathrm{T}_{\text{off}} = 3$, respectively. As an example, we choose the maximum tolerable detection delay as $50$ time units and if the attack cannot be detected within this period, we assume that the attack is missed. In Fig.~\ref{fig:miss_non_persistent}, we present the missed detection ratio versus average false alarm period for the proposed algorithms and the benchmark tests. As discussed in Sec.~\ref{sec:non_persistent}, against the non-persistent attacks, mainly the generalized Shewhart test is expected to perform well. That is, due to the off periods, even though the accumulated evidence supporting change may not become reliably high to declare an attack in Algorithm 1, the GLLR may take high values during the on periods. On the other hand, since the threshold of the generalized Shewhart test is chosen very high ($\phi = 10$) to reduce the false alarm level of Algorithm 2, the missed detection ratios in Algorithms 1 and 2 are almost the same for the small levels of average false alarm period, i.e., for the small test thresholds. However, for higher levels of average false alarm period, the missed detection ratio of Algorithm 2 significantly decreases compared to Algorithm 1 and the advantage of introducing the generalized Shewhart test in Algorithm 2 becomes visible in detecting the non-persistent stealthy attacks.

\subsection{Case 5: Persistent Stealthy Attack} \label{sec:per_st}

\begin{figure}
\center
  \includegraphics[width=78mm]{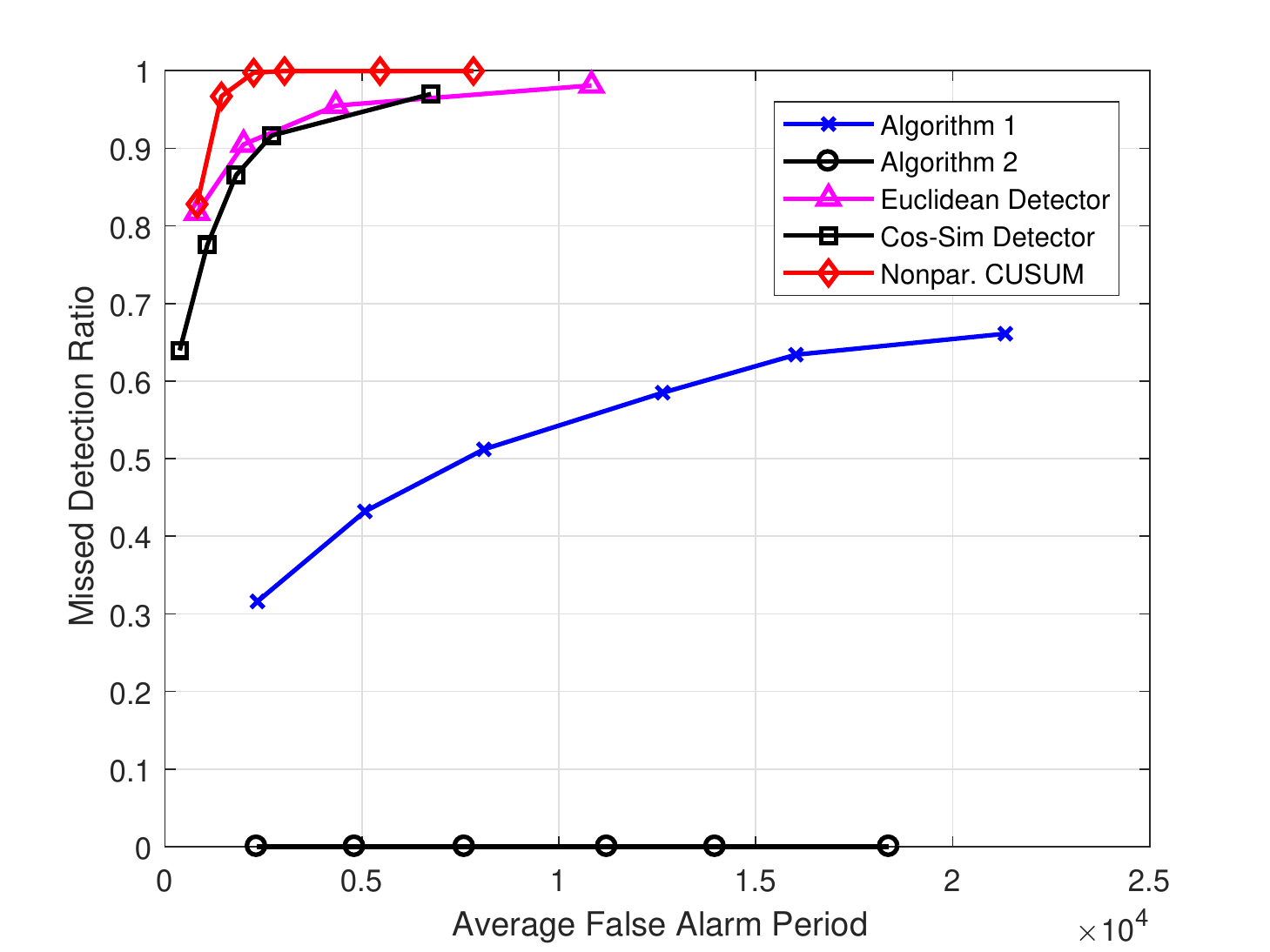}
\caption{Missed detection ratio vs. average false alarm period for the proposed detectors and the benchmark tests in case of a stealthy small-magnitude persistent attack, where the attack is assumed to be missed if it is not detected within $50$ time units.}
 \label{fig:miss_persistent}
\end{figure}

Although the considered lower bounds $\gamma$ and $\sigma^2$ on the attack magnitudes are already very small, an attacker may still perform a persistent stealthy attack using even lower attack magnitudes. Recall that we have previously showed in Figures \ref{fig:fdata_add_vs_mag} and \ref{fig:jamming_add_vs_mag} the advantage of Algorithm 2 over Algorithm 1 as the attack magnitudes get smaller for the FDI and jamming attacks, respectively. This time, we consider a hybrid attack with even smaller attack magnitudes where the magnitudes of FDI and jamming attacks are chosen as realizations of $\mathcal{U}[-0.005,0.005]$ and $\mathcal{U}[0.5\times10^{-4},10^{-4}]$, respectively. We present the missed detection ratio versus the average false alarm period curve for the proposed algorithms and the benchmark tests in Fig.~\ref{fig:miss_persistent}. We observe that Algorithm 2 significantly outperforms Algorithm 1 due to the introduced non-parametric chi-squared test in Algorithm 2. Since the attack magnitudes are very small, the proposed parametric tests become ineffective to detect such stealthy attacks. Note that although the non-parametric goodness-of-fit tests such as the chi-squared test becomes more successful in detecting such small-magnitude stealthy attacks, they in general lead to longer detection delays compared to the considered parametric tests since they usually require more samples for a reliable decision, mainly because they ignore all the prior knowledge about the post-attack case.

\subsection{Algorithm 1 vs. Countermeasures Against Stealthy Attacks}

\begin{figure}
\center
  \includegraphics[width=78mm]{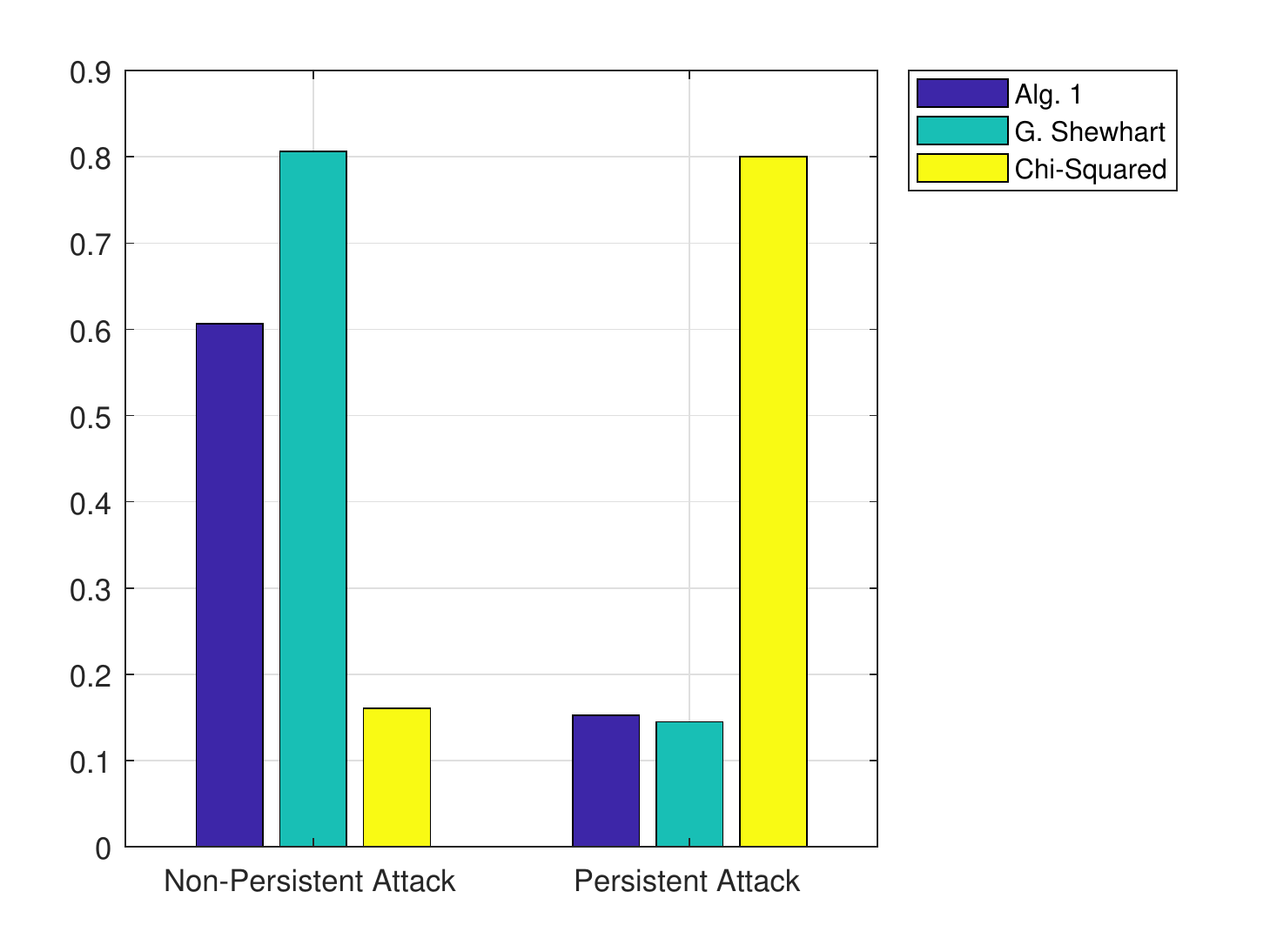}
\caption{Ratio of trials at which Algorithm 1, the generalized Shewhart test, and the sliding-window chi-squared test detect the stealthy attacks first (with the minimum delay), where the individual average false alarm periods of the algorithms are approximately $1.5\times10^{4}$.}
 \label{fig:bar_plot}
\end{figure}

With the purpose of illustrating the advantages of additional countermeasures employed in Algorithm 2 more clearly, Fig.~\ref{fig:bar_plot} shows a comparison between Algorithm 1 and the countermeasures in case of stealthy attacks described in Sec.~\ref{sec:nonper_st} and Sec.~\ref{sec:per_st}. Here, the individual average false alarm periods of Algorithm 1, the generalized Shewhart test, and the sliding-window chi-squared test are nearly equal to each other and for the non-persistent and persistent stealthy attacks, the figure shows the ratios over all trials at which each algorithm detects the attack first (with the minimum delay), where more than one test may simultaneously declare an attack with the minimum delay. Through the figure, we observe that the generalized Shewhart and the sliding-window chi-squared tests outperform Algorithm 1 in case of non-persistent and persistent stealthy attacks, respectively. Hence, together with the results obtained through Figures \ref{fig:miss_non_persistent} and \ref{fig:miss_persistent}, we can conclude that in case of stealthy attacks, the countermeasures improve the detection performance of Algorithm 2 compared to Algorithm 1.

\subsection{Recovered State Estimates}

\begin{figure}
\center
  \includegraphics[width=78mm]{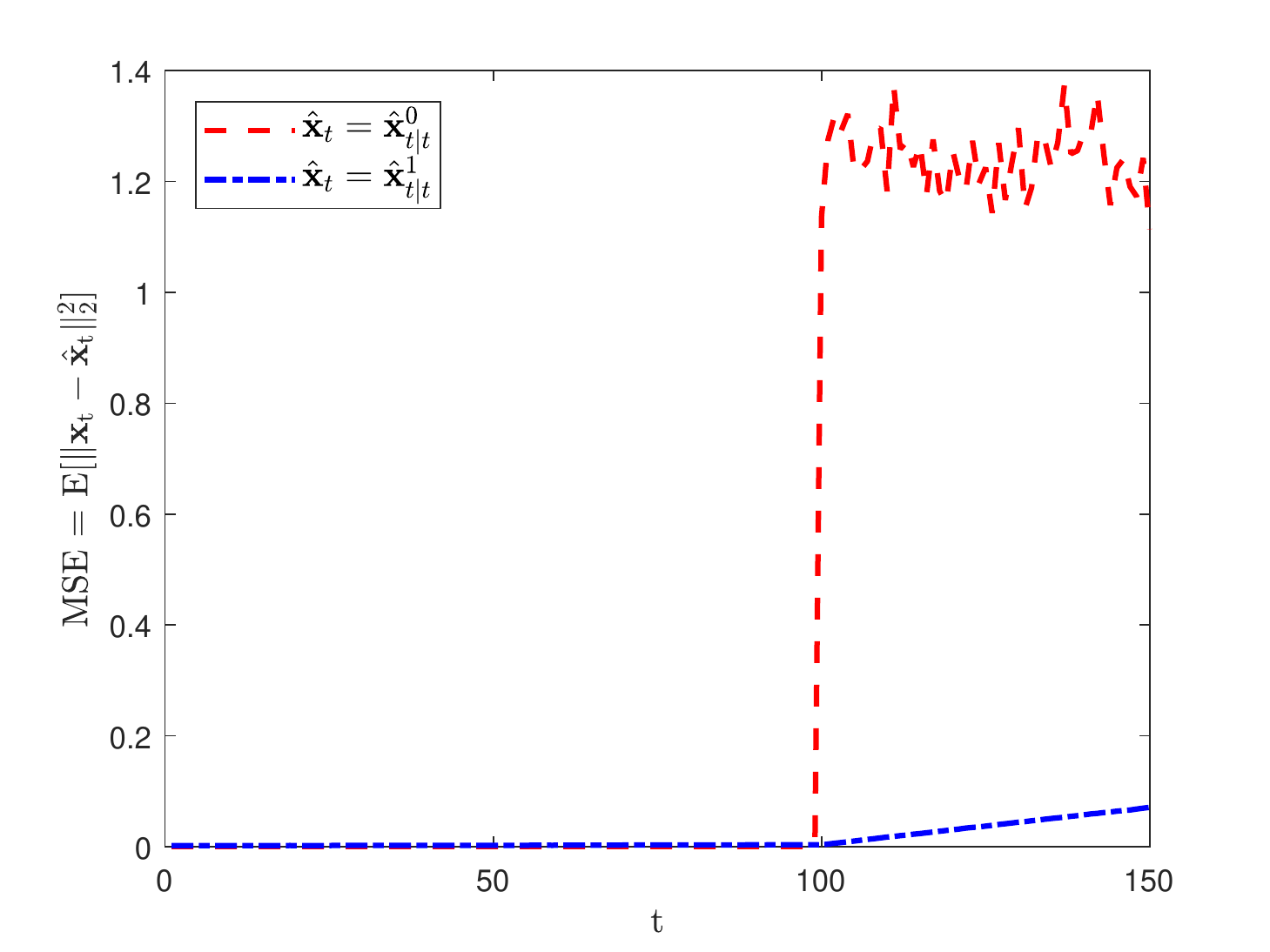}
\caption{MSE vs. time for the recovered ($\hat{\mathbf{x}}_{t|t}^1$) and non-recovered ($\hat{\mathbf{x}}_{t|t}^0$) state estimates in case of a hybrid attack.}
 \label{fig:mse_state}
\end{figure}

Fig.~\ref{fig:mse_state} presents the MSE versus time curve for the recovered, i.e., $\hat{\mathbf{x}}_{t|t}^1$, and the non-recovered, i.e., $\hat{\mathbf{x}}_{t|t}^0$, state estimates during the pre-change period, i.e., for $t < 100$, and the first $50$ time units after a hybrid FDI/jamming attack is launched to the system at $\tau = 100$. The FDI and jamming attacks are both of persistent nature as described in Sec.~\ref{sec:num_combined} and the attack magnitudes are realizations of $\mathcal{U}[-0.1,0.1]$ and $\mathcal{U}[1,2]$, respectively. Through the figure, we observe that the MSE of the recovered state estimates is significantly smaller than the MSE of the non-recovered state estimates. Further, we observe that the recovered state estimates slightly deviate from the actual system state $\mathbf{x}_t$ over the attacking period. This is due to the fact that the MLEs of the attack variables are computed based on the recovered state estimates (cf. \eqref{eq:a_hat_kt_v3_rrr} and \eqref{eq:sigma_hat_kt_v3_rrr}) and also the recovered state estimates are computed based on the MLEs of the attack variables (cf. \eqref{eq:meas_upd_fdata_alter}). Hence, the ML estimation errors accumulate over time during the attacking period. However, since the attacks can be quickly detected with the proposed real-time detection schemes, the deviation of the recovered state estimates is not expected to be significantly high at the detection time. Furthermore, recall that during the pre-attack period, whenever the decision statistic of Algorithm 1 reaches zero, the state estimates for the post-attack case are updated as being equal to the state estimates for the pre-attack case. Since the decision statistic frequently reaches zero during the pre-attack period, the ML estimation errors in computing the recovered state estimates do not accumulate in the pre-attack period.

\subsection{{Case 6: Topology Attack/Fault}}

{
Except the proposed nonparametric chi-squared test, the proposed methods are prone to the errors in the measurement matrix $\mathbf{H}$ due to either cyber-attacks or faults. This is because Algorithm 1 and the generalized Shewhart test are designed for a given $\mathbf{H}$ (see the hybrid attack model in \eqref{eq:meas_attacked}), whereas the chi-squared test does not depend on attack model assumptions.}

\begin{figure}
\center
  \includegraphics[width=78mm]{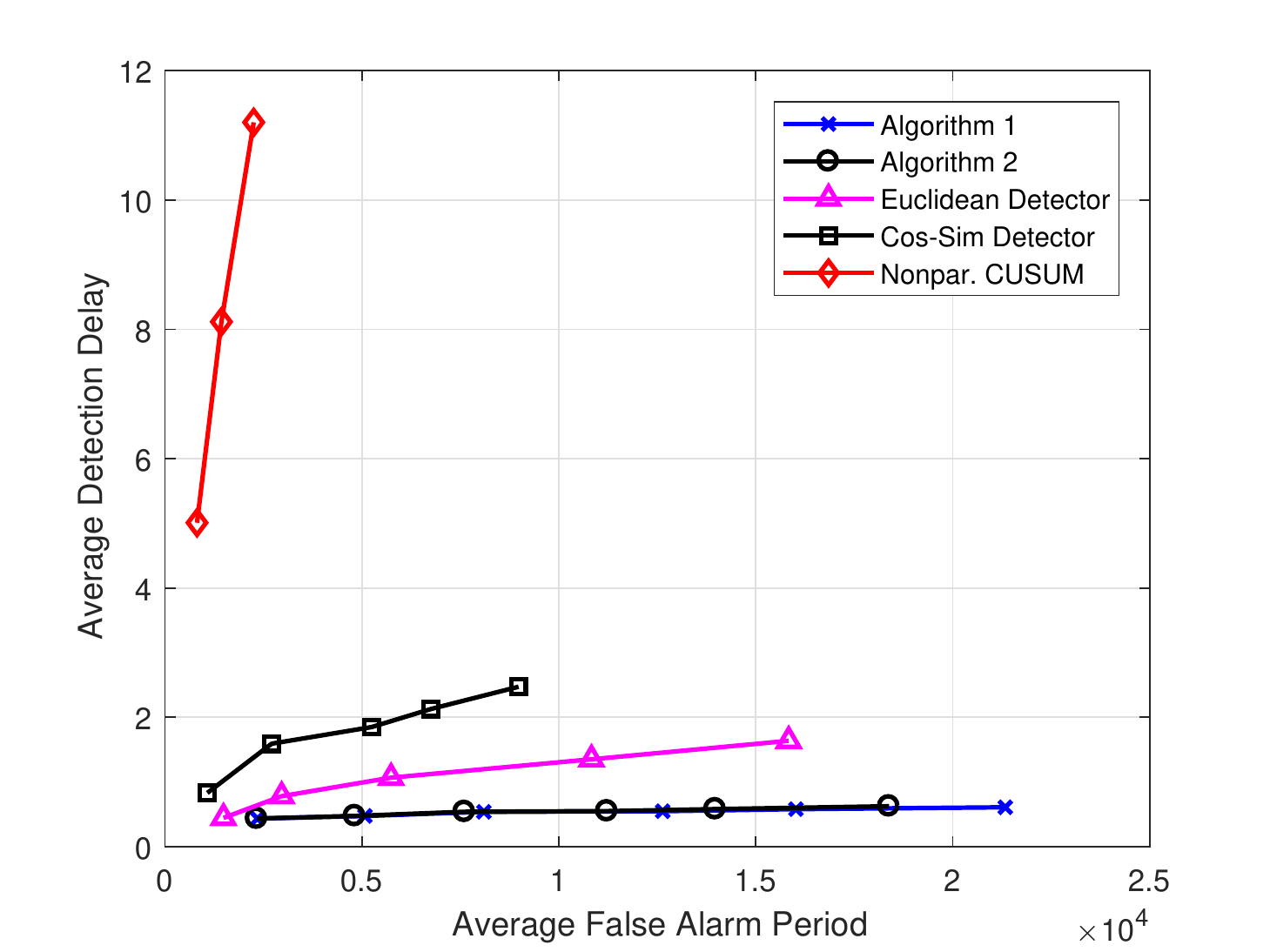}
\caption{{Average detection delay vs. average false alarm period for the proposed detectors and the benchmark tests in case of a network topology attack/fault.}}
 \label{fig:topology}
\end{figure}

{
On the other hand, the specific version of topology attack/failure in which some rows of $\mathbf{H}$ seem zero to the control center (although they are not) corresponds to DoS attacks, which is covered by the considered hybrid attack model (see Remark 1). For instance, if the link between two buses in a power grid breaks down due to an attack or fault, then the row in $\mathbf{H}$ corresponding to the power-flow measurement between these buses is changed accordingly such that the corresponding measurement signal becomes unavailable to the control center. Since the hybrid attack model covers DoS attacks as a special case, such topology attacks/faults can be detected by the proposed detectors. In Fig.~\ref{fig:topology}, we illustrate the performance of the proposed and the benchmark algorithms in detecting a line break between buses 9 and 10 in the IEEE-14 bus power system.}

\section{Conclusions} \label{sec:conc}

In this paper, we have studied the real-time detection of hybrid FDI/jamming attacks in the smart grid. {For a given network topology, we have modeled the smart grid as a linear dynamic system and employed Kalman filters for state estimation.} We have proposed an online CUSUM-based attack detection and estimation algorithm that is robust to unknown and time-varying attack parameters. We have also presented online estimates of the attack parameters in closed form and recovered state estimates in case of a cyber-attack. Furthermore, we have introduced and analyzed stealthy attacks against CUSUM-based detectors and specifically against the proposed algorithm, where the main aim of stealthy attacks is to prevent the detection or at least to increase the detection delays. {We have presented the generalized Shewhart test and the sliding-window chi-squared test as countermeasures against non-persistent and persistent stealthy attacks, respectively.} Through extensive simulations, we have illustrated that the proposed algorithms can timely and reliably detect {hybrid FDI/jamming attacks and stealthy attacks against CUSUM-based detectors, that correspond to} a significantly diverse range of potential cyber-attacks targeting the smart grid. {Moreover, the simulations illustrate the effectiveness of the proposed state recovery mechanism to mitigate the effects of cyber-attacks on the state estimation mechanism.}

{The proposed hybrid attack model does not cover network topology attacks as a special case. As a future work, the generalized state estimation mechanism \cite{Alsac98} can be considered where both the system state and the network topology are estimated based on power flow/injection measurements and measurements regarding the status of network switches and line breakers, and countermeasures can be developed against advanced topology attacks where attackers simultaneously perform hybrid FDI/jamming and network topology attacks.}

\appendix

\subsection{Proof of Proposition 1} \label{sec:proof_prop1}

Based on \eqref{eq:hyp_null} and \eqref{eq:hyp_alter}, $\beta_t$ in \eqref{eq:gen_CUSUM} can be written as follows:
\begin{gather} \nonumber
\beta_t = \frac{K \lambda}{2} \log(\sigma_w^2) + \frac{1}{2 \sigma_w^2} \sum_{k=1}^{K} \sum_{i=1}^{\lambda} (y_{k,t,i} - \mathbf{h}_k^\mathrm{T} \hat{\mathbf{x}}_t^0)^2
+ \sup_{\mathcal{S}_t^0,\mathcal{S}_t^f,\mathcal{S}_t^j,\mathcal{S}_t^{f,j}} \Bigg\{ \sup_{|a_{k,t}| \geq \gamma, \, k \, \in \, \mathcal{S}_t^f \, \cup \, \mathcal{S}_t^{f,j}} \, \sup_{\sigma_{k,t}^2 \geq \sigma^2, \, k \, \in \, \mathcal{S}_t^j \, \cup \, \mathcal{S}_t^{f,j}} \\ \nonumber
\bigg\{ \sum_{k \in \mathcal{S}_t^0} \sum_{i=1}^{\lambda} -\frac{1}{2} \log(\sigma_w^2) - \frac{1}{2 \sigma_w^2} (y_{k,t,i} - \mathbf{h}_k^\mathrm{T} \hat{\mathbf{x}}_t^1)^2
+ \sum_{k \in \mathcal{S}_t^f} \sum_{i=1}^{\lambda} -\frac{1}{2} \log(\sigma_w^2) - \frac{1}{2 \sigma_w^2} (y_{k,t,i} - \mathbf{h}_k^\mathrm{T} \hat{\mathbf{x}}_t^1 - a_{k,t})^2 \\ \nonumber
+ \sum_{k \in \mathcal{S}_t^j} \sum_{i=1}^{\lambda} -\frac{1}{2} \log(\sigma_w^2 + \sigma_{k,t}^2) - \frac{1}{2 (\sigma_w^2 + \sigma_{k,t}^2)} (y_{k,t,i} - \mathbf{h}_k^\mathrm{T} \hat{\mathbf{x}}_t^1)^2 \\ \label{eq:beta_t_v1}
+ \sum_{k \in \mathcal{S}_t^{f,j}} \sum_{i=1}^{\lambda} -\frac{1}{2} \log(\sigma_w^2 + \sigma_{k,t}^2) - \frac{1}{2 (\sigma_w^2 + \sigma_{k,t}^2)} (y_{k,t,i} - \mathbf{h}_k^\mathrm{T} \hat{\mathbf{x}}_t^1 - a_{k,t})^2  \bigg\} \Bigg\}.
\end{gather}

Let $e_{k,t,i} \triangleq y_{k,t,i} - \mathbf{h}_k^\mathrm{T} \hat{\mathbf{x}}_t^1$ and $\mathbf{e}_{k,t} \triangleq [e_{k,t,1}, e_{k,t,2}, \dots, e_{k,t,\lambda}]^\mathrm{T}$. Using the fact that taking supremum of a quantity is equivalent to taking infimum of the negative of the quantity, \eqref{eq:beta_t_v1} can be rewritten as
\begin{gather} \nonumber
\beta_t = \frac{K \lambda}{2} \log(\sigma_w^2) + \frac{1}{2 \sigma_w^2} \sum_{k=1}^{K} \sum_{i=1}^{\lambda} (y_{k,t,i} - \mathbf{h}_k^\mathrm{T} \hat{\mathbf{x}}_t^0)^2
- \frac{1}{2} \Bigg( \inf_{\mathcal{S}_t^0,\mathcal{S}_t^f,\mathcal{S}_t^j,\mathcal{S}_t^{f,j}} \bigg\{ \sum_{k \in \mathcal{S}_t^0} \underbrace{\lambda \log(\sigma_w^2) + \frac{\sum_{i=1}^{\lambda} e_{k,t,i}^2}{\sigma_w^2}}_{u^0(\mathbf{e}_{k,t})} \\ \nonumber
+ \sum_{k \in \mathcal{S}_t^f} \underbrace{\lambda \log(\sigma_w^2) + \frac{\inf_{|a_{k,t}| \geq \gamma} \big\{ \sum_{i=1}^{\lambda} (e_{k,t,i} - a_{k,t})^2 \big\}}{\sigma_w^2} }_{u^f(\mathbf{e}_{k,t})}
+ \sum_{k \in \mathcal{S}_t^j} \underbrace{ \inf_{\sigma_{k,t}^2 \geq \sigma^2} \Big\{ \lambda \log(\sigma_w^2 + \sigma_{k,t}^2) + \frac{ \sum_{i=1}^{\lambda} e_{k,t,i}^2}{(\sigma_w^2 + \sigma_{k,t}^2)} \Big\} }_{u^j(\mathbf{e}_{k,t})} \\ \label{eq:beta_tmp}
+ \sum_{k \in \mathcal{S}_t^{f,j}} \underbrace{ \inf_{\sigma_{k,t}^2 \geq \sigma^2} \inf_{|a_{k,t}| \geq \gamma} \Big\{ \lambda \log(\sigma_w^2 + \sigma_{k,t}^2) + \frac{\sum_{i=1}^{\lambda} (e_{k,t,i} - a_{k,t})^2}{(\sigma_w^2 + \sigma_{k,t}^2)} \Big\} }_{u^{f,j}(\mathbf{e}_{k,t})} \bigg\}  \Bigg) \\ \nonumber
=  \frac{K \lambda}{2} \log(\sigma_w^2) + \frac{1}{2 \sigma_w^2} \sum_{k=1}^{K} \sum_{i=1}^{\lambda} (y_{k,t,i} - \mathbf{h}_k^\mathrm{T} \hat{\mathbf{x}}_t^0)^2
- \frac{1}{2} \Bigg( \inf_{\mathcal{S}_t^0,\mathcal{S}_t^f,\mathcal{S}_t^j,\mathcal{S}_t^{f,j}} \bigg\{ \sum_{k \in \mathcal{S}_t^0} u^0(\mathbf{e}_{k,t}) + \sum_{k \in \mathcal{S}_t^f} u^f(\mathbf{e}_{k,t}) \\ \label{eq:beta_t_v2} \nonumber
+ \sum_{k \in \mathcal{S}_t^j} u^j(\mathbf{e}_{k,t}) + \sum_{k \in \mathcal{S}_t^{f,j}} u^{f,j}(\mathbf{e}_{k,t})  \bigg\}  \Bigg).
\end{gather}

The MLE estimates of $\mathcal{S}_t^0$, $\mathcal{S}_t^f$, $\mathcal{S}_t^j$, and $\mathcal{S}_t^{f,j}$ are then determined as follows:
\begin{align} \nonumber
&\hat{\mathcal{S}}_t^0 = \Big\{k: u^0(\mathbf{e}_{k,t}) \leq u^f(\mathbf{e}_{k,t}), \, u^0(\mathbf{e}_{k,t}) \leq u^j(\mathbf{e}_{k,t}), \, u^0(\mathbf{e}_{k,t}) \leq  u^{f,j}(\mathbf{e}_{k,t}), \, k = 1, 2, \dots, K \Big\} \\ \nonumber
&\hat{\mathcal{S}}_t^f = \Big\{ k: u^f(\mathbf{e}_{k,t}) < u^0(\mathbf{e}_{k,t}), \, u^f(\mathbf{e}_{k,t}) \leq u^j(\mathbf{e}_{k,t}), \, u^f(\mathbf{e}_{k,t}) \leq u^{f,j}(\mathbf{e}_{k,t}), \, k = 1, 2, \dots, K \Big\} \\ \nonumber
&\hat{\mathcal{S}}_t^j = \Big\{ k: u^j(\mathbf{e}_{k,t}) < u^0(\mathbf{e}_{k,t}), \, u^j(\mathbf{e}_{k,t}) < u^f(\mathbf{e}_{k,t}), \, u^j(\mathbf{e}_{k,t}) \leq u^{f,j}(\mathbf{e}_{k,t}), \, k = 1, 2, \dots, K \Big\} \\  \nonumber
&\hat{\mathcal{S}}_t^{f,j} = \Big\{ k: u^{f,j}(\mathbf{e}_{k,t}) < u^0(\mathbf{e}_{k,t}), \, u^{f,j}(\mathbf{e}_{k,t}) < u^f(\mathbf{e}_{k,t}), \, u^{f,j}(\mathbf{e}_{k,t}) < u^j(\mathbf{e}_{k,t}), \, k = 1, 2, \dots, K \Big\}.
\end{align}

Then, $\beta_t$ can be computed as
\begin{align} \nonumber
\beta_t &=  \frac{K \lambda}{2} \log(\sigma_w^2) + \frac{1}{2 \sigma_w^2} \sum_{k=1}^{K} \sum_{i=1}^{\lambda} (y_{k,t,i} - \mathbf{h}_k^\mathrm{T} \hat{\mathbf{x}}_t^0)^2 \\ \nonumber
&- \frac{1}{2} \bigg( \sum_{k \in \hat{\mathcal{S}}_t^0} u^0(\mathbf{e}_{k,t}) + \sum_{k \in \hat{\mathcal{S}}_t^f} u^f(\mathbf{e}_{k,t})
+ \sum_{k \in \hat{\mathcal{S}}_t^j} u^j(\mathbf{e}_{k,t}) + \sum_{k \in \hat{\mathcal{S}}_t^{f,j}} u^{f,j}(\mathbf{e}_{k,t}) \bigg),
\end{align}
where $u^0(\mathbf{e}_{k,t})$ is given by (cf. \eqref{eq:beta_tmp})
\begin{gather}\label{eq:u0} \nonumber
u^0(\mathbf{e}_{k,t}) =  \lambda \log(\sigma_w^2) + \frac{\sum_{i=1}^{\lambda} e_{k,t,i}^2}{\sigma_w^2}.
\end{gather}

Next, we determine $u^f(\mathbf{e}_{k,t})$, $u^j(\mathbf{e}_{k,t})$, $u^{f,j}(\mathbf{e}_{k,t})$, respectively and the MLEs of $a_{k,t}$ and $\sigma_{k,t}^2$. Firstly,
\begin{align} \nonumber
u^f(\mathbf{e}_{k,t}) &= \lambda \log(\sigma_w^2) + \frac{\inf_{|a_{k,t}| \geq \gamma} \big\{ \sum_{i=1}^{\lambda} (e_{k,t,i} - a_{k,t})^2 \big\}}{\sigma_w^2} \\  \nonumber
&= \begin{cases}
  \lambda \log(\sigma_w^2) + \frac{1}{\sigma_w^2} \sum_{i=1}^{\lambda} (e_{k,t,i} - \frac{1}{\lambda} \sum_{i=1}^{\lambda} e_{k,t,i})^2 , & \mbox{if } |\frac{1}{\lambda} \sum_{i=1}^{\lambda} e_{k,t,i}| \geq \gamma \\
  \lambda \log(\sigma_w^2) + \frac{1}{\sigma_w^2} \sum_{i=1}^{\lambda} (e_{k,t,i} - \gamma)^2, & \mbox{if } 0 \leq \frac{1}{\lambda} \sum_{i=1}^{\lambda} e_{k,t,i} < \gamma \\
  \lambda \log(\sigma_w^2) + \frac{1}{\sigma_w^2} \sum_{i=1}^{\lambda} (e_{k,t,i} + \gamma)^2, & \mbox{if } -\gamma < \frac{1}{\lambda} \sum_{i=1}^{\lambda} e_{k,t,i} < 0,
\end{cases}
\end{align}
where the MLE of $a_{k,t}, k \in \mathcal{S}_t^f$ is obtained as follows:
\begin{equation} \label{eq:a_hat_kt_v1}
    \hat{\mathrm{a}}_{k,t} =
    \begin{cases}
     \frac{1}{\lambda} \sum_{i=1}^{\lambda} e_{k,t,i} , & \text{if} ~~ |\frac{1}{\lambda} \sum_{i=1}^{\lambda} e_{k,t,i}| \geq \gamma, \, k \in \mathcal{S}_t^f \\
     \gamma , & \text{if} ~~ 0 \leq \frac{1}{\lambda} \sum_{i=1}^{\lambda} e_{k,t,i} < \gamma, \, k \in \mathcal{S}_t^f \\
     - \gamma , & \text{if} ~~ - \gamma < \frac{1}{\lambda} \sum_{i=1}^{\lambda} e_{k,t,i} < 0, \, k \in \mathcal{S}_t^f.
    \end{cases}
\end{equation}
Secondly,
\begin{align} \nonumber
u^j(\mathbf{e}_{k,t}) &= \inf_{\sigma_{k,t}^2 \geq \sigma^2} \Big\{ \lambda \log(\sigma_w^2 + \sigma_{k,t}^2) + \frac{ \sum_{i=1}^{\lambda} e_{k,t,i}^2}{\sigma_w^2 + \sigma_{k,t}^2} \Big\}.
\end{align}
We have $\frac{\partial u^j(\mathbf{e}_{k,t})}{\partial \sigma_{k,t}^2} = \frac{\lambda}{\sigma_w^2 + \sigma_{k,t}^2} - \frac{ \sum_{i=1}^{\lambda} e_{k,t,i}^2}{(\sigma_w^2 + \sigma_{k,t}^2)^2} = 0$ if $\sigma_{k,t}^2 = - \sigma_w^2 + \frac{1}{\lambda} \sum_{i=1}^{\lambda} e_{k,t,i}^2$. Moreover, for $\sigma_{k,t}^2 < - \sigma_w^2 + \frac{1}{\lambda} \sum_{i=1}^{\lambda} e_{k,t,i}^2$, $\frac{\partial u^j(\mathbf{e}_{k,t})}{\partial \sigma_{k,t}^2} < 0$ and for $\sigma_{k,t}^2 > - \sigma_w^2 + \frac{1}{\lambda} \sum_{i=1}^{\lambda} e_{k,t,i}^2$, $\frac{\partial u^j(\mathbf{e}_{k,t})}{\partial \sigma_{k,t}^2} > 0$. Hence, if $\frac{1}{\lambda} \sum_{i=1}^{\lambda} e_{k,t,i}^2 \geq \sigma_w^2 + \sigma^2$, $u^j(\mathbf{e}_{k,t})$ takes it minimum at $\sigma_{k,t}^2 = - \sigma_w^2 + \frac{1}{\lambda} \sum_{i=1}^{\lambda} e_{k,t,i}^2$. On the other hand, if $\frac{1}{\lambda} \sum_{i=1}^{\lambda} e_{k,t,i}^2 < \sigma_w^2 + \sigma^2$, $u^j(\mathbf{e}_{k,t})$ is monotone increasing function of $\sigma_{k,t}^2$ in the range of $\sigma_{k,t}^2 \geq \sigma^2$. Hence, $u^j(\mathbf{e}_{k,t})$ takes its minimum at $\sigma_{k,t}^2 = \sigma^2$. Then,
\begin{align}  \nonumber
u^j(\mathbf{e}_{k,t}) &= \begin{cases}
  \lambda \log(\frac{1}{\lambda} \sum_{i=1}^{\lambda} e_{k,t,i}^2) + \lambda, & \mbox{if } \frac{1}{\lambda} \sum_{i=1}^{\lambda} e_{k,t,i}^2 \geq \sigma_w^2 + \sigma^2 \\
  \lambda \log(\sigma_w^2 + \sigma^2) + \frac{1}{\sigma_w^2 + \sigma^2} \sum_{i=1}^{\lambda} e_{k,t,i}^2, & \mbox{if } \frac{1}{\lambda} \sum_{i=1}^{\lambda} e_{k,t,i}^2 < \sigma_w^2 + \sigma^2,
\end{cases}
\end{align}
where the MLE of $\sigma_{k,t}^2, k \in \mathcal{S}_t^j$ is obtained as follows:
\begin{equation} \label{eq:sigma_hat_kt_v1}
    \hat{\sigma}_{k,t}^2 =
    \begin{cases}
     - \sigma_w^2 + \frac{1}{\lambda} \sum_{i=1}^{\lambda} e_{k,t,i}^2 , & \text{if} ~~ \frac{1}{\lambda} \sum_{i=1}^{\lambda} e_{k,t,i}^2 \geq \sigma_w^2 + \sigma^2, \, k \in \mathcal{S}_t^j \\
     \sigma^2 , & \text{if} ~~ \frac{1}{\lambda} \sum_{i=1}^{\lambda} e_{k,t,i}^2 < \sigma_w^2 + \sigma^2, \, k \in \mathcal{S}_t^j.
    \end{cases}
\end{equation}
Further,
\begin{align} \nonumber
& u^{f,j}(\mathbf{e}_{k,t}) = \inf_{\sigma_{k,t}^2 \geq \sigma^2} \inf_{|a_{k,t}| \geq \gamma} \Big\{ \lambda \log(\sigma_w^2 + \sigma_{k,t}^2) + \frac{ \sum_{i=1}^{\lambda} (e_{k,t,i} - a_{k,t})^2}{(\sigma_w^2 + \sigma_{k,t}^2)} \Big\} \\ \nonumber
&= \inf_{\sigma_{k,t}^2 \geq \sigma^2} \Big\{ \lambda \log(\sigma_w^2 + \sigma_{k,t}^2) + \frac{ \inf_{|a_{k,t}| \geq \gamma} \big\{ \sum_{i=1}^{\lambda} (e_{k,t,i} - a_{k,t})^2 \big\}}{(\sigma_w^2 + \sigma_{k,t}^2)} \Big\} \\ \nonumber
&= \begin{cases}
       \inf_{\sigma_{k,t}^2 \geq \sigma^2} \Big\{ \lambda \log(\sigma_w^2 + \sigma_{k,t}^2) + \frac{1}{(\sigma_w^2 + \sigma_{k,t}^2)} \sum_{i=1}^{\lambda} (e_{k,t,i} - \frac{1}{\lambda} \sum_{i=1}^{\lambda} e_{k,t,i})^2 \Big\}, & \mbox{if } |\frac{1}{\lambda} \sum_{i=1}^{\lambda} e_{k,t,i}| \geq \gamma \\
       \inf_{\sigma_{k,t}^2 \geq \sigma^2} \Big\{ \lambda \log(\sigma_w^2 + \sigma_{k,t}^2) + \frac{1}{(\sigma_w^2 + \sigma_{k,t}^2)} \sum_{i=1}^{\lambda} (e_{k,t,i} - \gamma)^2 \Big\}, & \mbox{if } 0 \leq \frac{1}{\lambda} \sum_{i=1}^{\lambda} e_{k,t,i} < \gamma \\
       \inf_{\sigma_{k,t}^2 \geq \sigma^2} \Big\{ \lambda \log(\sigma_w^2 + \sigma_{k,t}^2) + \frac{1}{(\sigma_w^2 + \sigma_{k,t}^2)}  \sum_{i=1}^{\lambda} (e_{k,t,i} + \gamma)^2 \Big\}, & \mbox{if } -\gamma < \frac{1}{\lambda} \sum_{i=1}^{\lambda} e_{k,t,i} < 0
   \end{cases} \\ \nonumber
&= \begin{cases}
     \lambda \log(\frac{1}{\lambda} \sum_{i=1}^{\lambda} (e_{k,t,i} - \frac{1}{\lambda} \sum_{i=1}^{\lambda} e_{k,t,i})^2) + \lambda, & \mbox{if } |\frac{1}{\lambda} \sum_{i=1}^{\lambda} e_{k,t,i}| \geq \gamma \mbox{ and } \frac{1}{\lambda} \sum_{i=1}^{\lambda} (e_{k,t,i} - \frac{1}{\lambda} \sum_{i=1}^{\lambda} e_{k,t,i})^2 \geq \sigma_w^2 + \sigma^2 \\
     \lambda \log(\sigma_w^2 + \sigma^2) + \frac{1}{\sigma_w^2 + \sigma^2} \sum_{i=1}^{\lambda} (e_{k,t,i} - \frac{1}{\lambda} \sum_{i=1}^{\lambda} e_{k,t,i})^2, & \mbox{if } |\frac{1}{\lambda} \sum_{i=1}^{\lambda} e_{k,t,i}| \geq \gamma \mbox{ and } \frac{1}{\lambda} \sum_{i=1}^{\lambda} (e_{k,t,i} - \frac{1}{\lambda} \sum_{i=1}^{\lambda} e_{k,t,i})^2 < \sigma_w^2 + \sigma^2 \\
     \lambda \log(\frac{1}{\lambda} \sum_{i=1}^{\lambda} (e_{k,t,i} - \gamma)^2) + \lambda, & \mbox{if } 0 \leq \frac{1}{\lambda} \sum_{i=1}^{\lambda} e_{k,t,i} < \gamma \mbox{ and } \frac{1}{\lambda} \sum_{i=1}^{\lambda} (e_{k,t,i} - \gamma)^2 \geq \sigma_w^2 + \sigma^2 \\
     \lambda \log(\sigma_w^2 + \sigma^2) + \frac{1}{\sigma_w^2 + \sigma^2} \sum_{i=1}^{\lambda} (e_{k,t,i} - \gamma)^2, & \mbox{if } 0 \leq \frac{1}{\lambda} \sum_{i=1}^{\lambda} e_{k,t,i} < \gamma \mbox{ and } \frac{1}{\lambda} \sum_{i=1}^{\lambda} (e_{k,t,i} - \gamma)^2 < \sigma_w^2 + \sigma^2 \\
     \lambda \log(\frac{1}{\lambda} \sum_{i=1}^{\lambda} (e_{k,t,i} + \gamma)^2) + \lambda, & \mbox{if } -\gamma < \frac{1}{\lambda} \sum_{i=1}^{\lambda} e_{k,t,i} < 0 \mbox{ and } \frac{1}{\lambda} \sum_{i=1}^{\lambda} (e_{k,t,i} + \gamma)^2 \geq \sigma_w^2 + \sigma^2 \\
     \lambda \log(\sigma_w^2 + \sigma^2) + \frac{1}{\sigma_w^2 + \sigma^2} \sum_{i=1}^{\lambda} (e_{k,t,i} + \gamma)^2, & \mbox{if } -\gamma < \frac{1}{\lambda} \sum_{i=1}^{\lambda} e_{k,t,i} < 0 \mbox{ and } \frac{1}{\lambda} \sum_{i=1}^{\lambda} (e_{k,t,i} + \gamma)^2 < \sigma_w^2 + \sigma^2,
   \end{cases}
\end{align}
where the MLE of $a_{k,t}, k \in \mathcal{S}_t^{f,j}$ is obtained as
\begin{equation} \label{eq:a_hat_kt_v2}
    \hat{\mathrm{a}}_{k,t} =
    \begin{cases}
     \frac{1}{\lambda} \sum_{i=1}^{\lambda} e_{k,t,i} , & \text{if} ~~ |\frac{1}{\lambda} \sum_{i=1}^{\lambda} e_{k,t,i}| \geq \gamma, \, k \in \mathcal{S}_t^{f,j} \\
     \gamma , & \text{if} ~~ 0 \leq \frac{1}{\lambda} \sum_{i=1}^{\lambda} e_{k,t,i} < \gamma, \, k \in \mathcal{S}_t^{f,j} \\
     - \gamma , & \text{if} ~~ - \gamma < \frac{1}{\lambda} \sum_{i=1}^{\lambda} e_{k,t,i} < 0, \, k \in \mathcal{S}_t^{f,j},
    \end{cases}
\end{equation}
and the MLE of $\sigma_{k,t}^2, k \in \mathcal{S}_t^{f,j}$ is obtained as follows:
\begin{equation} \label{eq:sigma_hat_kt_v2}
    \hat{\sigma}_{k,t}^2 =
    \begin{cases}
     - \sigma_w^2 + \frac{1}{\lambda} \sum_{i=1}^{\lambda} (e_{k,t,i} - \frac{1}{\lambda} \sum_{i=1}^{\lambda} e_{k,t,i})^2, & \mbox{if } |\frac{1}{\lambda} \sum_{i=1}^{\lambda} e_{k,t,i}| \geq \gamma \mbox{ and } \frac{1}{\lambda} \sum_{i=1}^{\lambda} (e_{k,t,i} - \frac{1}{\lambda} \sum_{i=1}^{\lambda} e_{k,t,i})^2 \geq \sigma_w^2 + \sigma^2 \\
     \sigma^2, & \mbox{if } |\frac{1}{\lambda} \sum_{i=1}^{\lambda} e_{k,t,i}| \geq \gamma \mbox{ and } \frac{1}{\lambda} \sum_{i=1}^{\lambda} (e_{k,t,i} - \frac{1}{\lambda} \sum_{i=1}^{\lambda} e_{k,t,i})^2 < \sigma_w^2 + \sigma^2 \\
     - \sigma_w^2 + \frac{1}{\lambda} \sum_{i=1}^{\lambda} (e_{k,t,i} - \gamma)^2, & \mbox{if } 0 \leq \frac{1}{\lambda} \sum_{i=1}^{\lambda} e_{k,t,i} < \gamma \mbox{ and } \frac{1}{\lambda} \sum_{i=1}^{\lambda} (e_{k,t,i} - \gamma)^2 \geq \sigma_w^2 + \sigma^2 \\
     \sigma^2, & \mbox{if } 0 \leq \frac{1}{\lambda} \sum_{i=1}^{\lambda} e_{k,t,i} < \gamma \mbox{ and } \frac{1}{\lambda} \sum_{i=1}^{\lambda} (e_{k,t,i} - \gamma)^2 < \sigma_w^2 + \sigma^2 \\
     - \sigma_w^2 + \frac{1}{\lambda} \sum_{i=1}^{\lambda} (e_{k,t,i} + \gamma)^2, & \mbox{if } -\gamma < \frac{1}{\lambda} \sum_{i=1}^{\lambda} e_{k,t,i} < 0 \mbox{ and } \frac{1}{\lambda} \sum_{i=1}^{\lambda} (e_{k,t,i} + \gamma)^2 \geq \sigma_w^2 + \sigma^2 \\
     \sigma^2, & \mbox{if } -\gamma < \frac{1}{\lambda} \sum_{i=1}^{\lambda} e_{k,t,i} < 0 \mbox{ and } \frac{1}{\lambda} \sum_{i=1}^{\lambda} (e_{k,t,i} + \gamma)^2 < \sigma_w^2 + \sigma^2.
    \end{cases}
\end{equation}

Using \eqref{eq:a_hat_kt_v1}, \eqref{eq:a_hat_kt_v2}, and the MLEs of $\mathcal{S}_t^0$, $\mathcal{S}_t^f$, $\mathcal{S}_t^j$, and $\mathcal{S}_t^{f,j}$, the MLE of $a_{k,t}, k \in \{1,2,\dots,K\}$ is then determined as follows:
\begin{equation} \nonumber
    \hat{\mathrm{a}}_{k,t} =
    \begin{cases}
     \frac{1}{\lambda} \sum_{i=1}^{\lambda} e_{k,t,i}, & \text{if} ~~ |\frac{1}{\lambda} \sum_{i=1}^{\lambda} e_{k,t,i}| \geq \gamma \mbox{ and } k \in \hat{\mathcal{S}}_t^f \cup \hat{\mathcal{S}}_t^{f,j} \\
     \gamma , & \text{if} ~~ 0 \leq \frac{1}{\lambda} \sum_{i=1}^{\lambda} e_{k,t,i} < \gamma \mbox{ and } k \in \hat{\mathcal{S}}_t^f \cup \hat{\mathcal{S}}_t^{f,j} \\
     - \gamma , & \text{if} ~~ -\gamma < \frac{1}{\lambda} \sum_{i=1}^{\lambda} e_{k,t,i} < 0 \mbox{ and } k \in \hat{\mathcal{S}}_t^f \cup \hat{\mathcal{S}}_t^{f,j} \\
     0, & \text{if} ~~ k \in \hat{\mathcal{S}}_t^0 \cup \hat{\mathcal{S}}_t^j.
    \end{cases}
\end{equation}

Furthermore, using \eqref{eq:sigma_hat_kt_v1}, \eqref{eq:sigma_hat_kt_v2}, and  the MLEs of $\mathcal{S}_t^0$, $\mathcal{S}_t^f$, $\mathcal{S}_t^j$, and $\mathcal{S}_t^{f,j}$, the MLE of $\sigma_{k,t}^2, k \in \{1,2,\dots,K\}$ is obtained as follows:
\begin{align} \nonumber
   & \hat{\sigma}_{k,t}^2 = \\ \nonumber
   & \begin{cases}
     - \sigma_w^2 + \frac{1}{\lambda} \sum_{i=1}^{\lambda} e_{k,t,i}^2 , & \text{if} ~~ \frac{1}{\lambda} \sum_{i=1}^{\lambda} e_{k,t,i}^2 \geq \sigma_w^2 + \sigma^2 \mbox{ and } k \in \hat{\mathcal{S}}_t^j  \\
     \sigma^2 , & \text{if} ~~ \frac{1}{\lambda} \sum_{i=1}^{\lambda} e_{k,t,i}^2 < \sigma_w^2 + \sigma^2 \mbox{ and } k \in \hat{\mathcal{S}}_t^j \\
     - \sigma_w^2 + \frac{1}{\lambda} \sum_{i=1}^{\lambda} (e_{k,t,i} - \frac{1}{\lambda} \sum_{i=1}^{\lambda} e_{k,t,i})^2, & \mbox{if } |\frac{1}{\lambda} \sum_{i=1}^{\lambda} e_{k,t,i}| \geq \gamma \mbox{ and } \frac{1}{\lambda} \sum_{i=1}^{\lambda} (e_{k,t,i} - \frac{1}{\lambda} \sum_{i=1}^{\lambda} e_{k,t,i})^2 \geq \sigma_w^2 + \sigma^2 \mbox{ and } k \in \hat{\mathcal{S}}_t^{f,j} \\
     \sigma^2, & \mbox{if } |\frac{1}{\lambda} \sum_{i=1}^{\lambda} e_{k,t,i}| \geq \gamma \mbox{ and } \frac{1}{\lambda} \sum_{i=1}^{\lambda} (e_{k,t,i} - \frac{1}{\lambda} \sum_{i=1}^{\lambda} e_{k,t,i})^2 < \sigma_w^2 + \sigma^2 \mbox{ and } k \in \hat{\mathcal{S}}_t^{f,j} \\
     - \sigma_w^2 + \frac{1}{\lambda} \sum_{i=1}^{\lambda} (e_{k,t,i} - \gamma)^2, & \mbox{if } 0 \leq \frac{1}{\lambda} \sum_{i=1}^{\lambda} e_{k,t,i} < \gamma \mbox{ and } \frac{1}{\lambda} \sum_{i=1}^{\lambda} (e_{k,t,i} - \gamma)^2 \geq \sigma_w^2 + \sigma^2 \mbox{ and } k \in \hat{\mathcal{S}}_t^{f,j} \\
     \sigma^2, & \mbox{if } 0 \leq \frac{1}{\lambda} \sum_{i=1}^{\lambda} e_{k,t,i} < \gamma \mbox{ and } \frac{1}{\lambda} \sum_{i=1}^{\lambda} (e_{k,t,i} - \gamma)^2 < \sigma_w^2 + \sigma^2 \mbox{ and } k \in \hat{\mathcal{S}}_t^{f,j} \\
     - \sigma_w^2 + \frac{1}{\lambda} \sum_{i=1}^{\lambda} (e_{k,t,i} + \gamma)^2, & \mbox{if } -\gamma < \frac{1}{\lambda} \sum_{i=1}^{\lambda} e_{k,t,i} < 0 \mbox{ and } \frac{1}{\lambda} \sum_{i=1}^{\lambda} (e_{k,t,i} + \gamma)^2 \geq \sigma_w^2 + \sigma^2 \mbox{ and } k \in \hat{\mathcal{S}}_t^{f,j} \\
     \sigma^2, & \mbox{if } -\gamma < \frac{1}{\lambda} \sum_{i=1}^{\lambda} e_{k,t,i} < 0 \mbox{ and } \frac{1}{\lambda} \sum_{i=1}^{\lambda} (e_{k,t,i} + \gamma)^2 < \sigma_w^2 + \sigma^2 \mbox{ and } k \in \hat{\mathcal{S}}_t^{f,j} \\
     0, & \mbox{if } k \in \hat{\mathcal{S}}_t^0 \cup \hat{\mathcal{S}}_t^f.
    \end{cases}
\end{align}

Finally, defining $\delta_{k,t} \triangleq \sum_{i=1}^{\lambda} e_{k,t,i}$, $\zeta_{k,t} \triangleq \sum_{i=1}^{\lambda} e_{k,t,i}^2$, $\varrho_{k,t} \triangleq \sum_{i=1}^{\lambda} (e_{k,t,i} + \gamma)^2$, and $\varpi_{k,t} \triangleq \sum_{i=1}^{\lambda} (e_{k,t,i} - \gamma)^2$, $\forall k \in \{1,2,\dots,K\}$ and $\forall t > 0$, we obtain the simplified expressions as given in Proposition 1.

\qed

\bibliographystyle{IEEEtran}
\bibliography{det_refs,refs}

\begin{thebibliography}{10}
\providecommand{\url}[1]{#1}
\csname url@samestyle\endcsname
\providecommand{\newblock}{\relax}
\providecommand{\bibinfo}[2]{#2}
\providecommand{\BIBentrySTDinterwordspacing}{\spaceskip=0pt\relax}
\providecommand{\BIBentryALTinterwordstretchfactor}{4}
\providecommand{\BIBentryALTinterwordspacing}{\spaceskip=\fontdimen2\font plus
\BIBentryALTinterwordstretchfactor\fontdimen3\font minus
  \fontdimen4\font\relax}
\providecommand{\BIBforeignlanguage}[2]{{%
\expandafter\ifx\csname l@#1\endcsname\relax
\typeout{** WARNING: IEEEtran.bst: No hyphenation pattern has been}%
\typeout{** loaded for the language `#1'. Using the pattern for}%
\typeout{** the default language instead.}%
\else
\language=\csname l@#1\endcsname
\fi
#2}}
\providecommand{\BIBdecl}{\relax}
\BIBdecl

\bibitem{HHe16}
H.~He and J.~Yan, ``Cyber-physical attacks and defences in the smart grid: a
  survey,'' \emph{IET Cyber-Physical Systems: Theory Applications}, vol.~1,
  no.~1, pp. 13--27, 2016.

\bibitem{Liang16}
G.~Liang, J.~Zhao, F.~Luo, S.~Weller, and Z.~Y. Dong, ``A review of false data
  injection attacks against modern power systems,'' \emph{IEEE Transactions on
  Smart Grid}, vol.~PP, no.~99, pp. 1--1, 2016.

\bibitem{wang2013cyber}
W.~Wang and Z.~Lu, ``Cyber security in the smart grid: Survey and challenges,''
  \emph{Computer Networks}, vol.~57, no.~5, pp. 1344--1371, 2013.

\bibitem{yan2012survey}
Y.~Yan, Y.~Qian, H.~Sharif, and D.~Tipper, ``A survey on cyber security for
  smart grid communications,'' \emph{IEEE Communications Surveys \& Tutorials},
  2012.

\bibitem{Xie10}
L.~Xie, Y.~Mo, and B.~Sinopoli, ``False data injection attacks in electricity
  markets,'' in \emph{2010 First IEEE International Conference on Smart Grid
  Communications}, Oct 2010, pp. 226--231.

\bibitem{moslemi2017}
R.~Moslemi, A.~Mesbahi, and J.~Mohammadpour, ``Design of robust profitable
  false data injection attacks in multi-settlement electricity markets,''
  \emph{IET Generation, Transmission \& Distribution}, 2017.

\bibitem{ayar2017}
M.~Ayar, S.~Obuz, R.~D. Trevizan, A.~S. Bretas, and H.~A. Latchman, ``A
  distributed control approach for enhancing smart grid transient stability and
  resilience,'' \emph{IEEE Transactions on Smart Grid}, vol.~8, no.~6, pp.
  3035--3044, 2017.

\bibitem{GLiang17}
G.~Liang, S.~R. Weller, J.~Zhao, F.~Luo, and Z.~Y. Dong, ``The 2015 ukraine
  blackout: Implications for false data injection attacks,'' \emph{IEEE
  Transactions on Power Systems}, vol.~32, no.~4, pp. 3317--3318, July 2017.

\bibitem{Amin09}
S.~Amin, A.~A. C{\'a}rdenas, and S.~S. Sastry, \emph{Safe and Secure Networked
  Control Systems under Denial-of-Service Attacks}.\hskip 1em plus 0.5em minus
  0.4em\relax Berlin, Heidelberg: Springer Berlin Heidelberg, 2009, pp. 31--45.

\bibitem{Liu09}
Y.~Liu, P.~Ning, and M.~K. Reiter, ``False data injection attacks against state
  estimation in electric power grids,'' in \emph{Proceedings of the 16th ACM
  Conference on Computer and Communications Security}, ser. CCS '09.\hskip 1em
  plus 0.5em minus 0.4em\relax New York, NY, USA: ACM, 2009, pp. 21--32.

\bibitem{Tan17}
S.~Tan, D.~De, W.~Z. Song, J.~Yang, and S.~K. Das, ``Survey of security
  advances in smart grid: A data driven approach,'' \emph{IEEE Communications
  Surveys Tutorials}, vol.~19, no.~1, pp. 397--422, Firstquarter 2017.

\bibitem{moslemi2017fast}
R.~Moslemi, A.~Mesbahi, and J.~M. Velni, ``A fast, decentralized covariance
  selection-based approach to detect cyber attacks in smart grids,'' \emph{IEEE
  Transactions on Smart Grid}, 2017.

\bibitem{Sargolzaei17}
A.~Sargolzaei, K.~Yen, M.~Abdelghani, A.~Abbaspour, and S.~Sargolzaei,
  ``Generalized attack model for networked control systems, evaluation of
  control methods,'' \emph{Intelligent Control and Automation}, vol.~08, pp.
  164--174, 2017.

\bibitem{YLi15}
Y.~Li, L.~Shi, P.~Cheng, J.~Chen, and D.~E. Quevedo, ``Jamming attacks on
  remote state estimation in cyber-physical systems: A game-theoretic
  approach,'' \emph{IEEE Transactions on Automatic Control}, vol.~60, no.~10,
  pp. 2831--2836, Oct 2015.

\bibitem{Deka15}
D.~Deka, R.~Baldick, and S.~Vishwanath, ``Optimal data attacks on power grids:
  Leveraging detection measurement jamming,'' in \emph{2015 IEEE International
  Conference on Smart Grid Communications (SmartGridComm)}, Nov 2015, pp.
  392--397.

\bibitem{Abur04}
A.~Abur and A.~Gomez-Exposito, \emph{Power System State Estimation: Theory and
  Implementation}, 01 2004, vol.~24.

\bibitem{Manandhar14}
K.~Manandhar, X.~Cao, F.~Hu, and Y.~Liu, ``Detection of faults and attacks
  including false data injection attack in smart grid using kalman filter,''
  \emph{IEEE Transactions on Control of Network Systems}, vol.~1, no.~4, pp.
  370--379, Dec 2014.

\bibitem{Brumback87}
B.~Brumback and M.~Srinath, ``A chi-square test for fault-detection in kalman
  filters,'' \emph{IEEE Transactions on Automatic Control}, vol.~32, no.~6, pp.
  552--554, Jun 1987.

\bibitem{Rawat15}
D.~B. Rawat and C.~Bajracharya, ``Detection of false data injection attacks in
  smart grid communication systems,'' \emph{IEEE Signal Processing Letters},
  vol.~22, no.~10, pp. 1652--1656, Oct 2015.

\bibitem{Esmalifalak17}
M.~Esmalifalak, L.~Liu, N.~Nguyen, R.~Zheng, and Z.~Han, ``Detecting stealthy
  false data injection using machine learning in smart grid,'' \emph{IEEE
  Systems Journal}, vol.~11, no.~3, pp. 1644--1652, Sept 2017.

\bibitem{Ozay16}
M.~Ozay, I.~Esnaola, F.~T.~Y. Vural, S.~R. Kulkarni, and H.~V. Poor, ``Machine
  learning methods for attack detection in the smart grid,'' \emph{IEEE
  Transactions on Neural Networks and Learning Systems}, vol.~27, no.~8, pp.
  1773--1786, Aug 2016.

\bibitem{bretas2017}
A.~S. Bretas, N.~G. Bretas, B.~Carvalho, E.~Baeyens, and P.~P. Khargonekar,
  ``Smart grids cyber-physical security as a malicious data attack: An
  innovation approach,'' \emph{Electric Power Systems Research}, vol. 149, pp.
  210--219, 2017.

\bibitem{zhao2017robust}
J.~Zhao, M.~Netto, and L.~Mili, ``A robust iterated extended kalman filter for
  power system dynamic state estimation,'' \emph{IEEE Transactions on Power
  Systems}, vol.~32, no.~4, pp. 3205--3216, 2017.

\bibitem{zhao2017attack}
J.~Zhao, L.~Mili, and A.~Abdelhadi, ``Robust dynamic state estimator to
  outliers and cyber attacks,'' in \emph{Power \& Energy Society General
  Meeting, 2017 IEEE}.\hskip 1em plus 0.5em minus 0.4em\relax IEEE, 2017, pp.
  1--5.

\bibitem{gandhi2010robust}
M.~A. Gandhi and L.~Mili, ``Robust kalman filter based on a generalized
  maximum-likelihood-type estimator,'' \emph{IEEE Transactions on Signal
  Processing}, vol.~58, no.~5, pp. 2509--2520, 2010.

\bibitem{Li_15}
S.~Li, Y.~Yilmaz, and X.~Wang, ``Quickest detection of false data injection
  attack in wide-area smart grids,'' \emph{IEEE Transactions on Smart Grid},
  vol.~6, no.~6, pp. 2725--2735, Nov 2015.

\bibitem{Huang16}
Y.~Huang, J.~Tang, Y.~Cheng, H.~Li, K.~A. Campbell, and Z.~Han, ``Real-time
  detection of false data injection in smart grid networks: An adaptive cusum
  method and analysis,'' \emph{IEEE Systems Journal}, vol.~10, no.~2, pp.
  532--543, June 2016.

\bibitem{Necip18}
M.~N. Kurt, Y.~Yilmaz, and X.~Wang, ``Distributed quickest detection of
  cyber-attacks in smart grid,'' \emph{IEEE Transactions on Information
  Forensics and Security}, vol.~13, no.~8, pp. 2015--2030, Aug 2018.

\bibitem{yang2016false}
Q.~Yang, L.~Chang, and W.~Yu, ``On false data injection attacks against kalman
  filtering in power system dynamic state estimation,'' \emph{Security and
  Communication Networks}, vol.~9, no.~9, pp. 833--849, 2016.

\bibitem{sun2013anomaly}
B.~Sun, X.~Shan, K.~Wu, and Y.~Xiao, ``Anomaly detection based secure
  in-network aggregation for wireless sensor networks,'' \emph{IEEE Systems
  Journal}, vol.~7, no.~1, pp. 13--25, 2013.

\bibitem{Cui12}
S.~Cui, Z.~Han, S.~Kar, T.~T. Kim, H.~V. Poor, and A.~Tajer, ``Coordinated
  data-injection attack and detection in the smart grid: A detailed look at
  enriching detection solutions,'' \emph{IEEE Signal Processing Magazine},
  vol.~29, no.~5, pp. 106--115, Sept 2012.

\bibitem{Zhao17}
J.~Zhao, G.~Zhang, M.~L. Scala, Z.~Y. Dong, C.~Chen, and J.~Wang, ``Short-term
  state forecasting-aided method for detection of smart grid general false data
  injection attacks,'' \emph{IEEE Transactions on Smart Grid}, vol.~8, no.~4,
  pp. 1580--1590, July 2017.

\bibitem{JGao15}
J.~Gao, S.~A. Vorobyov, H.~Jiang, and H.~V. Poor, ``Worst-case jamming on mimo
  gaussian channels,'' \emph{IEEE Transactions on Signal Processing}, vol.~63,
  no.~21, pp. 5821--5836, Nov 2015.

\bibitem{Gezici16}
S.~Gezici, S.~Bayram, M.~N. Kurt, and M.~R. Gholami, ``Optimal jammer placement
  in wireless localization systems,'' \emph{IEEE Transactions on Signal
  Processing}, vol.~64, no.~17, pp. 4534--4549, Sept 2016.

\bibitem{Kay93}
S.~M. Kay, \emph{Fundamentals of Statistical Signal Processing: Estimation
  Theory}.\hskip 1em plus 0.5em minus 0.4em\relax Upper Saddle River, NJ, USA:
  Prentice-Hall, Inc., 1993.

\bibitem{Kalman_60}
R.~E. Kalman, ``A new approach to linear filtering and prediction problems,''
  \emph{Transactions of the ASME--Journal of Basic Engineering}, vol.~82, no.
  Series D, pp. 35--45, 1960.

\bibitem{Poor08}
H.~V. Poor and O.~Hadjiliadis, \emph{{Quickest Detection}}.\hskip 1em plus
  0.5em minus 0.4em\relax Cambridge University Press, 2008.

\bibitem{Basseville93}
M.~Basseville and I.~V. Nikiforov, \emph{Detection of Abrupt Changes: Theory
  and Application}.\hskip 1em plus 0.5em minus 0.4em\relax Upper Saddle River,
  NJ, USA: Prentice-Hall, Inc., 1993.

\bibitem{Veeravalli14}
V.~V. Veeravalli and T.~Banerjee, ``Chapter 6 - quickest change detection,'' in
  \emph{Academic Press Library in Signal Processing: Volume 3 Array and
  Statistical Signal Processing}, ser. Academic Press Library in Signal
  Processing, R.~C. Abdelhak M.~Zoubir, Mats~Viberg and S.~Theodoridis,
  Eds.\hskip 1em plus 0.5em minus 0.4em\relax Elsevier, 2014, vol.~3, pp. 209
  -- 255.

\bibitem{Lorden_71}
G.~Lorden, ``Procedures for reacting to a change in distribution,'' \emph{Ann.
  Math. Statist.}, vol.~42, no.~6, pp. 1897--1908, 1971.

\bibitem{Moustakides_86}
G.~V. Moustakides, ``Optimal stopping times for detecting changes in
  distributions,'' \emph{Ann. Statist.}, vol.~14, no.~4, pp. 1379--1387, 1986.

\bibitem{Moustakides14}
------, ``Multiple optimality properties of the shewhart test,''
  \emph{Sequential Analysis}, vol.~33, no.~3, pp. 318--344, 2014.

\bibitem{Zimmerman11}
R.~Zimmerman, C.~Murillo-Sanchez, and R.~Thomas, ``Matpower: Steady-state
  operations, planning, and analysis tools for power systems research and
  education,'' \emph{IEEE Transactions on Power Systems}, vol.~26, no.~1, pp.
  12--19, Feb 2011.

\bibitem{Alsac98}
O.~Alsac, N.~Vempati, B.~Stott, and A.~Monticelli, ``Generalized state
  estimation,'' \emph{IEEE Transactions on power systems}, vol.~13, no.~3, pp.
  1069--1075, 1998.

\end{thebibliography}

\end{document}